\documentclass{article}

\usepackage{blindtext}
\usepackage{hyperref}
\hypersetup{
    colorlinks=true,
    linkcolor=blue,
    filecolor=magenta,      
    urlcolor=cyan,
    pdftitle={Overleaf Example},
    pdfpagemode=FullScreen,
    }
\usepackage{amsfonts}
\usepackage{cite}
\usepackage{graphicx}
\usepackage{dcolumn}
\usepackage{bm}
\usepackage{amsmath}
\usepackage{amssymb}
\usepackage{hhline}
\usepackage[normalem]{ulem}
\usepackage{authblk}

\usepackage[numbers,sort,compress]{natbib}

\usepackage{color}
\date{}

\title{Polariton lattices as binarized neuromorphic networks}
\author{Evgeny Sedov$^{1,2,3}$\thanks{Email: evgeny\_sedov@mail.ru}, Alexey Kavokin$^{1,3,4}$}
\affil{$^1$Spin-Optics laboratory, St. Petersburg State University, St. Petersburg 198504, Russia}
\affil{$^2$Stoletov Vladimir State University, Vladimir 600000, Russia}
\affil{$^3$School of Science, Westlake University, 600 Dunyu Road, Hangzhou 310030, Zhejiang Province, China}
\affil{$^4$Abrikosov Center for Theoretical Physics, Moscow Institute of Physics and Technology, Dolgoprudnyi, Moscow Region 141701 Russia}

\begin{document}

\maketitle

\begin{abstract}
We introduce a novel neuromorphic network architecture based on a lattice of exciton-polariton condensates, intricately interconnected and energized through non-resonant optical pumping.
The network employs a binary framework, where each neuron, facilitated by the spatial coherence of pairwise coupled condensates, performs binary operations.
This coherence, emerging from the ballistic propagation of polaritons, ensures efficient, network-wide communication.
The binary neuron switching mechanism, driven by the nonlinear repulsion through the excitonic component of polaritons, offers computational efficiency and scalability advantages over continuous weight neural networks.
Our network enables parallel processing, enhancing computational speed compared to sequential or pulse-coded binary systems.
{
The system's performance was evaluated using diverse datasets, including the MNIST dataset for image recognition and the Speech Commands dataset for voice recognition tasks.
In both scenarios, the proposed system demonstrates the potential to outperform existing polaritonic neuromorphic systems.
For image recognition, this is evidenced by an impressive predicted classification accuracy of up to 97.5\%.
In voice recognition, the system achieved a classification accuracy of about 68\% for the ten-class subset, surpassing the performance of conventional benchmark, the Hidden Markov Model with Gaussian Mixture Model.}
\end{abstract}

%SSSSSSSSSSSSSSSSS
%EEEEEEEEEEEEEEEEE
%CCCCCCCCCCCCCCCCC
\section*{Introduction}

The rapid development of artificial neural networks and applied artificial intelligence is predominantly aimed at the efficient processing of large data sets and pattern recognition~\cite{APRes7011312}.
Traditional approaches, however, are increasingly facing constraints in terms of computational speed and energy efficiency, particularly in hardware implementations of these networks~\cite{FrontNeur16453X}.
These constraints have spurred interest in neuromorphic systems, where hardware mimics the structure and function of the human brain. The exploration of novel materials and mechanisms is crucial for the development of efficient neuromorphic systems~\cite{NatMat15517}.

A particularly promising direction in this research area is the exploitation of exciton-polariton interactions within specially designed semiconductor microcavities in the strong light-matter coupling regime~\cite{Nanolett203506,PhysRevApplied18024028}. %~\cite{Nanolett203506,PhysRevApplied18024028,OptMatExpr132674}.
Exciton-polaritons are quasiparticles emerging from the coupling of photons and excitons~\cite{kavokinBook2017}. They possess dual light and matter properties, enabling strong optical nonlinearity and picosecond scale reaction times.
These characteristics  enable the development of polariton based high-speed neuromorphic systems with high efficiency~\cite{RevModPhys85299}.

The term ``polariton neuron'' was first introduced in Ref.~\cite{PhysRevLett101016402}, devoted to planar waveguide structures  that translate polariton coherence over extended distances.
This research laid the groundwork for using polariton neurons to construct binary logic gates in semiconductor microcavities, serving as a sort of precursor for neuromorphic computing. 
Much later, the reservoir computing scheme has emerged as a key approach for the development of polariton-based neural networks~\cite{PhysRevApplied11064029, PhysRevApplied13064074}. 
This technique employs a network with fixed, random connections, simplifying the architecture as compared to traditional neural networks.
Historically, quantum computing has been considered the ``holy grail'' of polaritonics, particularly in terms of the applicability of its outcomes.
This is why the concept of reservoir computing has been highly regarded, as it extends its utility from classical to quantum computing domains~\cite{PhysRevLett123260404}. 
Complementing this quantum focus, recent advances of polaritonics might bring important applications in classical information processing, particularly in pattern recognition.

Reflecting the latter statement, the authors of Ref.~\cite{Nanolett203506} leveraged the polariton properties to achieve a 93\% level in recognizing handwritten digits in the Mixed National Institute of Standards and Technology dataset (MNIST dataset)~\cite{lecun1998gradient,lecun2010mnist}, that is a benchmark in pattern recognition tasks. Complementing this, Ref.~\cite{PhysRevApplied18024028} reported not only a 96.2\% classification accuracy on the same dataset in the experiment but also demonstrated the efficiency of backpropagation training in their exciton-polariton-based neuromorphic hardware.
This efficiency in processing complex patterns, however, often comes with increased computational and memory demands in traditional neuromorphic networks~\cite{Nature604255}, which necessitates either a reduction in input resolution or more complex processing architectures for feasible operation times.

In contrast, the introduction of binarized neural networks, which streamline the network by utilizing two-level activations or weights and performing simple binary operations, marks a significant advancement in the field. These networks are distinguished by their enhanced speed and energy efficiency, with only a minimal trade-off in inference accuracy~\cite{NanoLett213715}.
By efficiently using memory to store binary rather than continuous data and simplifying computational demands, binarized networks offer a compelling alternative to conventional networks with continuous activation functions.
In the context of high-speed neuromorphic systems, binary networks have shown a significant progress.
Recent advancements in the area of neuromorphic binarized polariton networks have showcased their remarkable capabilities.
As shown in Ref.~\cite{NanoLett213715}, this approach, involving input encoding using nonresonant picosecond laser pulses to excite localized condensation sites, each representing binary inputs with distinct pulse energies, has led to significant achievements in pattern recognition. 
Notably, the system achieves approximately 96\% classification accuracy on the MNIST dataset, even in a noisy experimental environment, using a single-hidden-layer network.
This level of accuracy, attained through binary operations, is particularly impressive given the challenges of the experimental setup.

In scenarios where information is encoded through individual or paired pulses, there's a marked tendency to opt for a sequential processing approach, employing a single transformation gate for each pulse set.
However, this approach presents challenges in terms of operational speed within these binary systems.
A tool for overcoming the issue of parallelizing input uploading in neuromorphic binarized polariton networks has been proposed in a recent study~\cite{PhysRevApplied16024045}.
Their method involves spatial encoding of input information, with all input pulses designed to arrive at the network simultaneously.
By enabling parallel input encoding, this method effectively addresses the operational speed limitations inherent in the sequential approach.

The next natural step towards parallelizing neuronal triggering, in conjunction with ensuring the interaction among individual neurons, is the use of spatial lattices of neurons.
Lattices of mesoscopic coherent condensates of exciton polariton have evolved as a sophisticated extension of the principles observed in chains and lattices of ultracold atoms~\cite{RevModPhys80885,Nature41539,PhysRevA86023829,PhysRevA89033828,NatPhys18657} 
harnessing the unique properties of light-matter interactions to delve into new frontiers of quantum simulations and condensed matter physics~\cite{CompRenPhys17934,NatRevPhys4435,NJPhys19125008,NatMater161120}. 
Advancing beyond their predecessors, they offer greater control and versatility, operating at higher temperatures and allowing for more dynamic configurations, thereby enhancing their practicality in exploring complex quantum phenomena~\cite{NatRevPhys4435}.

Various techniques have been employed to manipulate the spatial potential for trapping and arranging polaritons in a microcavity plane~\cite{CompRenPhys17934}.
Among these techniques are etching lattices of coupled micropillars from planar microcavities~\cite{PhysRevLett120097401}, variation of the thickness of the cavity layer ~\cite{PhysRevB74155311}, deposition of metallic films onto the surface of the microcavity~\cite{NatPhys7681}.

An alternative approach to creating polariton lattices in a microcavity plane is by using regular spatial patterns of the pumping light.
In this geometry, each condensate in a lattice is created by a separate non-resonant optical pump beam, while spatial coherence between the condensates across the lattice is provided by the exchange of ballistically propagating polaritons~\cite{Optica8106}.
Besides replenishing the polariton condensate state, the pump also facilitates their trapping, contributing to the formation of an effective complex potential for the trap.
Depending on the combination of kinetic properties of the polaritons and the gain-loss balance, either dissipative trapping can occur, where the condensates are predominantly localized within the pump spot ~\cite{Optica8106,NJPhys19125008,PhysRevLett124207402}  (phase locking regime), or they are localized in the minima of the real part of the effective potential created by the pumping light~\cite{PhysRevB92035305,PhysRevB97235303}.
The complex effective potential is formed by the repulsive reservoir of hot incoherent excitons, which are excited by the pump light.
Spatial light modulators offer an advantage of high spatial resolution and the ability to control the intensity and distribution of the reservoir with great accuracy, enabling precise manipulation of the locations of polariton condensates and their coupling strengths within the lattice.
Additionally, these optically induced potentials can be effectively combined with stationary potentials, offering even more versatility of control over polariton condensates~\cite{PhysRevResearch3023167,PhysRevB97195149,SciRep134607,CommunPhys559}.

In this manuscript, we present a neuromorphic network architecture employing lattices of exciton-polariton condensates, interconnected and energized through non-resonant optical pumping.
This design capitalizes on benefits of a binary framework, wherein each neuron, aided by the spatial coherence of coupled condensates, executes binary operations.
The lattice structure facilitates parallel uploading and processing of signals, enhancing neuron-neuron interactions.
The effectiveness of our system was evaluated using the MNIST dataset.
It demonstrated promising results compared to current neuromorphic systems.
Additionally, we have developed a technique for input signal densing that notably improves the system performance, that achieves an accuracy rate of 97.5\%, surpassing that of existing polariton-based neuromorphic systems.
{
To further validate our model's robustness and versatility, we extended our testing to the Speech Commands dataset~\cite{warden2018speechcommandsdatasetlimitedvocabulary,speechcommandsv2}, which includes audio clips for voice command recognition.
Our system's performance, considerably exceeding that of both the linear classifier and the advanced Hidden Markov Model with Gaussian Mixture Model (HMM-GMM)~\cite{IEEETrSP134406,SIG0042008}, confirms its potential for effective application in complex speech recognition tasks.}

%FFFFFFFFFFFFFFFFF
%IIIIIIIIIIIIIIIII
%GGGGGGGGGGGGGGGGG
\begin{figure*}[tb!]
\begin{center}
\includegraphics[width=\linewidth]{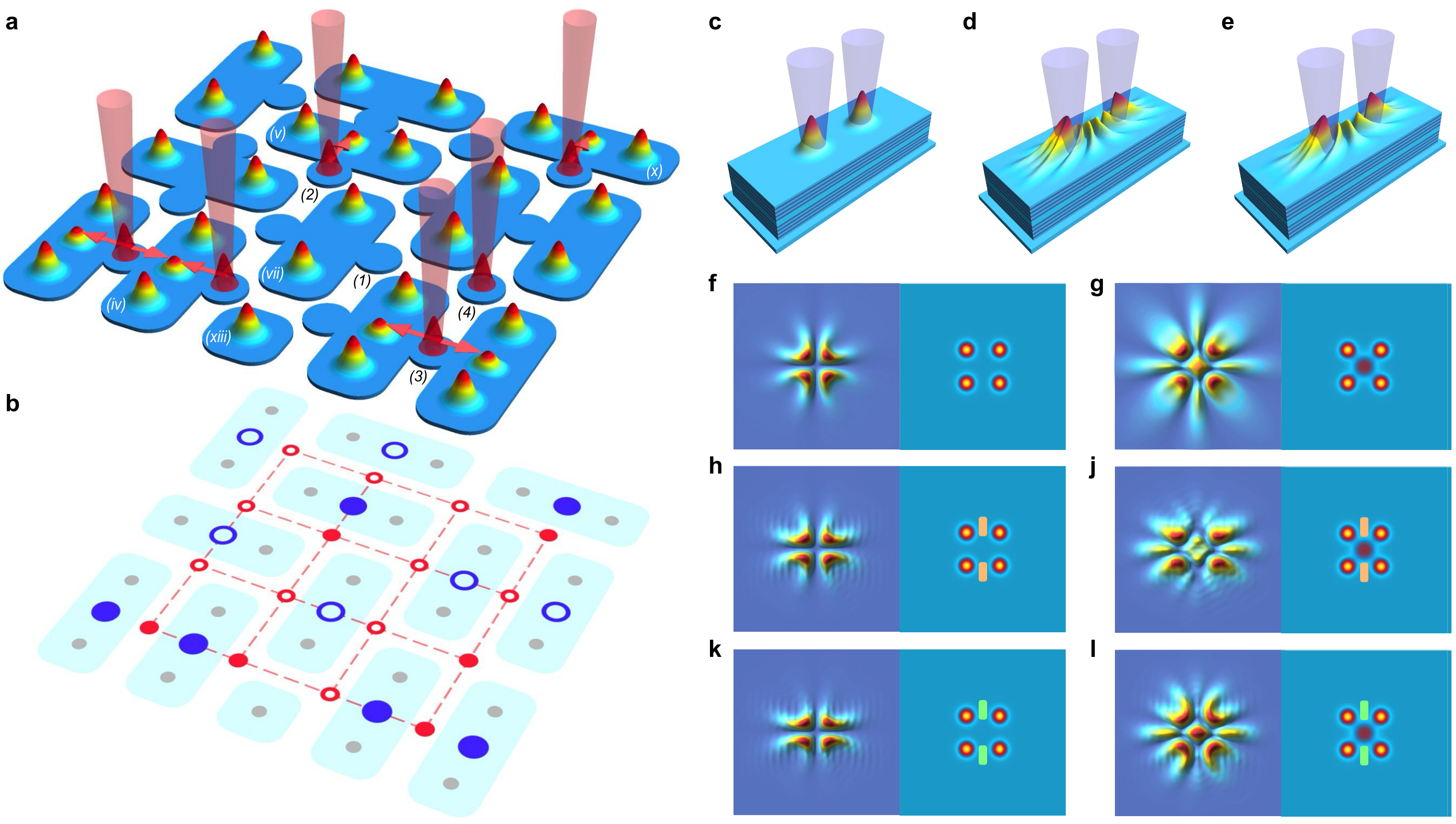}
\end{center}
\caption{
(a) A diagram depicting the possible experimental implementation of a lattice of pairwise coupled polariton condensates with optically controlled connections.
Red cones symbolize input signal beams, and red arrows indicate the connections influenced by these beams.
The dyads are numbered with Roman numerals to correspond with the numbering in Fig.~\ref{FIG_ManyDyads}.
(b) A streamlined illustration of the envisaged structure, with gray, red, and blue circles denoting condensate lattice nodes, input, and output optical signals, respectively.
Empty (filled) circles correspond to the absence (presence) of the signals.
(c--e) Illustration of polariton dyad excitation in a planar microcavity, showing (c) profiles of two non-resonant optical pump spots for dyad excitation, (d,e) the condensates in the dyad with even (OFF) and odd (ON) interference patterns.
(f-l) Depiction of the excitation of two adjacent dyads in OFF (f,h,k) and ON (g,j,l) states.
Each pair of panels shows polariton density distribution (left) and pump intensity profiles, including the potential barrier (right).
Switching between OFF and ON states is achieved using a signal optical pump beam equidistant from the four nodes.
Panels (f,g) illustrate no separation between the dyads,
(h,j) and (k,l) show the dyads separated by real (orange rectangle) and imaginary (blue rectangle) potential barriers, respectively. }
\label{FIG_Scheme1}
\end{figure*}

%FFFFFFFFFFFFFFFFF
%IIIIIIIIIIIIIIIII
%GGGGGGGGGGGGGGGGG
\begin{figure*}[tb!]
\begin{center}
\includegraphics[width=\linewidth]{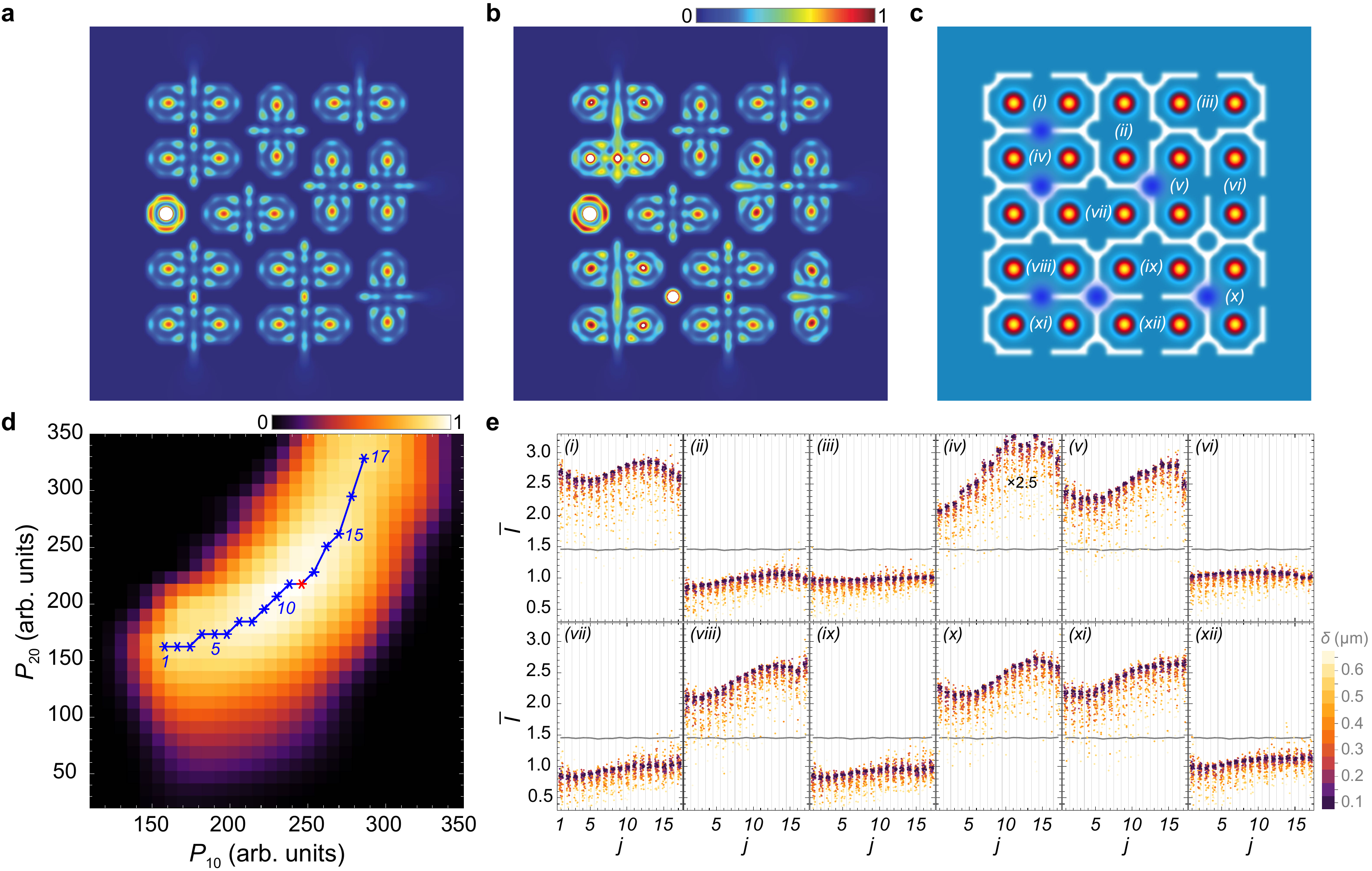}
\end{center}
\caption{
Depiction of excitation of a lattice of dyads of geometry, schematically shown in Figs.~\ref{FIG_Scheme1}(a) and~\ref{FIG_Scheme1}(b), in the absence (a) and in the presence (b) of control signals.
The panels show time-integrated spatial distribution of the polariton density.
Regions with densities higher than the range covered by the color scale are indicated in white.
(c) The profiles of the non-resonant pump spots (main color scheme), the trapping potential profile for isolating dyads (white) and the control beam profiles (blue) for toggling dyads between ON and OFF states.
The indices \textit{(i)} to \textit{(xii)} enumerate neurons in (a) and (b).
(d) The dependence of distinguishability, $\Delta \tilde{I}$, of the OFF and ON signals on the intensities of the pump pulses for condensates in dyads, $P_{10}$, and intensities of signal pulses, $P_{20}$.
Values of $\Delta \tilde{I}<0$ are colored in black.
Pulse durations are taken as $w_{\tau1} = 5$~ps and $w_{\tau2} = 8$~ps, respectively. 
Definition of the distinguishability, $\Delta \tilde{I}$, is given in the text.
Star markers enumerated from 1 to 17 indicate maximal $\Delta \tilde{I}$ at given~$P_{10}$.
The red star indicates the parameters used for (a) and (b).
(e) Variation of the relative intensity of output signals, $\bar{I}$, in each neuron from \textit{(i)} to \textit{(xii)} at the pump intensities $(P_{10},P_{20})$ corresponding to points $j=1,2,... 17$ in (d).
In \textit{(iv)}, the dependence should be multiplied by 2.5.
Each dot corresponds to a separate numerical experiment, in which positions of the pump pulses, that excite condensates in dyads across the lattice, deviate randomly from their owing positions in the range of distances from $-\delta$ to $+\delta$. 
Dots of different colors correspond to different deviations $\delta$. 
Gray lines used as references indicate the average of the minimal relative intensity of the neuron in the ON state and the maximal relative intensity of the neuron in the OFF state in the absence of deviations of the pump pulses. 
}
\label{FIG_ManyDyads}
\end{figure*}

%SSSSSSSSSSSSSSSSS
%EEEEEEEEEEEEEEEEE
%CCCCCCCCCCCCCCCCC
\section*{Results}

%SSSSSSSSSSSSSSSSS
%SSSSSSSSSSSSSSSSS
%EEEEEEEEEEEEEEEEE
%CCCCCCCCCCCCCCCCC
\subsection*{The physical background of a polariton neuromorphic network}

We consider a lattice of pairwise coupled polariton condensates --- \textit{polariton dyads}, the connections between which can be manipulated through an external optical impact.
A schematic of such a lattice as well as its possible implementations are depicted in Fig.~\ref{FIG_Scheme1}(a,b).
In developing our structure, we drew inspiration from a series of publications~\cite{PhysRevLett124207402,NJPhys19125008,Optica8106} dedicated to establishing and controlling spatial coherence in lattices of polariton condensates.

A polariton dyad represents a pair of polariton condensates excited by localized optical beams in a plane of a microcavity, separated by a distance $d$ from each other, see Figs.~\ref{FIG_Scheme1}(c--e).
Excitation is performed in the non-resonant regime, where the energy of an excitation beam is considerably (tens of meV) higher than the polariton energy~\cite{Nature443409,ACSPhot71163}.
Such a pump creates a reservoir of incoherent high-energy excitons (see Fig.~\ref{FIG_Scheme1}(c)), which in turn feeds the polariton condensate.
This process is facilitated by the stimulated scattering of quasiparticles, accompanied by a reduction of their energy.
In addition to its role as a feeder, the exciton reservoir acts as a potential barrier for the condensate due to a strong repulsive polariton-exciton interaction.

Polaritons, whose lifetime may be reduced by tailoring the quality factor of the microcavity, are primarily localized near the pump spot, engaging in the earlier mentioned dissipative trapping regime.
However, in each specific condensate, polaritons demonstrate the radial ballistic expansion, driven by their repulsion from the potential barrier created by the reservoir.
The coupling and coherence build up within the condensate dyad are facilitated by fast ballistically propagating polaritons, whose propagation distances exceed the condensates' separation distance~$d$.

%FFFFFFFFFFFFFFFFF
%IIIIIIIIIIIIIIIII
%GGGGGGGGGGGGGGGGG
\begin{figure*}[tb!]
\begin{center}
\includegraphics[width=\linewidth]{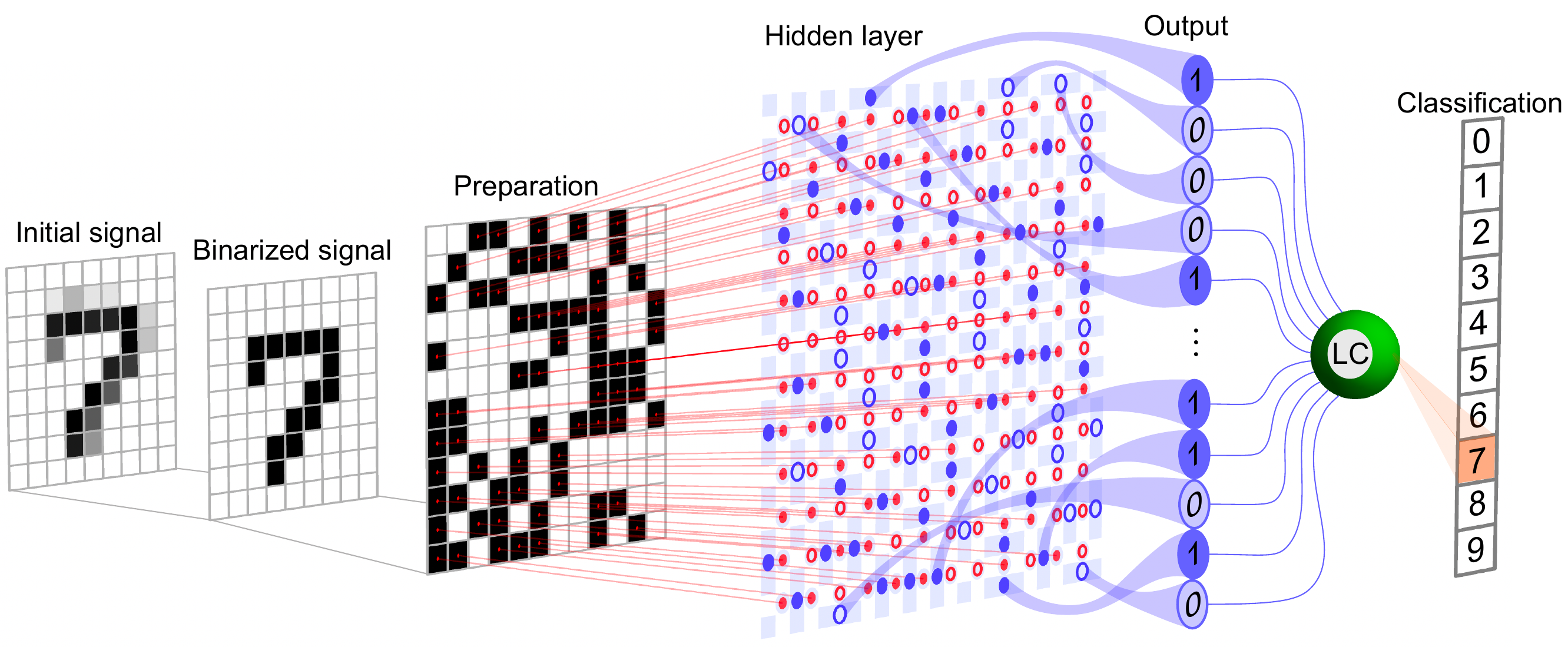}
\end{center}
\caption{Conceptual diagram of a binarized neural network based on a lattice of pairwise coupled polariton condensates.
The initial signal originates from a %28$\times$28 pixel 
grayscale image from the MNIST dataset, which is binarized and projected onto a $n _{\text{in}} \times n _{\text{in}}$ transformation lattice.
This lattice serves as a pattern for the input optical signal.
This signal then activates neurons within the hidden layer, generating the resultant optical output signal.
Subsequently, the output is processed via a linear classifier~(LC).}
\label{FIG_Scheme2}
\end{figure*}

The spatial coherence of the condensates manifests itself through the emergence of interference patterns within the space between the condensates.
If both condensates are pumped with equal strengths, these patterns are symmetrical and can possess an even or odd number of interference fringes, with either a minimum or maximum intensity at the midpoint between them, cf. Figs.~\ref{FIG_Scheme1}(d) and~\ref{FIG_Scheme1}(e).
The coupling between the condensates can be enhanced by adjusting the shape of the pump, as discussed in Ref.~\cite{PhysRevB106245304}.

In a planar microcavity, the parity of the interference pattern is controlled by the distance $d$ and the optical pumping power.
The latter, in particular, defines the height of the potential barrier induced by the reservoir.
To alter the parity, one can make an adjustment to the potential landscape within the dyad.
Namely, an additional exciton reservoir can be excited by a laser beam, whose power and spatial profile can be arbitrarily chosen within the bounds of the interference pattern, as suggested in Ref.~\cite{PhysRevLett124207402}.
The left panels of Figs.~\ref{FIG_Scheme1}(f,g) illustrate two adjacent polariton dyads, numerically simulated by solving the complex generalized Gross-Pitaevskii equation (see Methods for details).
This configuration can be excited by non-resonant optical beams forming a unit cell of a square lattice, as schematically shown in the right panel of Fig.~\ref{FIG_Scheme1}(f).
These figures demonstrate the capability to simultaneously switch both dyads between OFF and ON states using an additional single optical beam centered in the cell, see the right panel of Fig.~\ref{FIG_Scheme1}(g).
In simulations, the pump power and the condensates' separation distance are optimized to minimize the number of interference fringes between the condensates.
Thus, in the OFF state for both dyads, in the absence of the additional control beam (Fig.~\ref{FIG_Scheme1}(f)), the number of fringes is reduced to zero, resulting in a minimum between the condensates.
Whilst the control beam is present (Fig.~\ref{FIG_Scheme1}(g)), both dyads switch to the ON state, characterized by a single interference fringe that forms a maximum intensity between the condensates.
The photoluminescence signal coming from the space between the condensates can be experimentally detected, allowing for the unambiguous distinction between the OFF and ON states.

As observed in the geometry presented in Figs.~\ref{FIG_Scheme1}(f,g), the condensates are interconnected not only within each dyad but also with other condensates belonging to an adjacent dyad.
In such a configuration, it's more appropriate to refer to a \textit{tetrad}, involving all condensates excited within an elementary cell.
Moreover, in the limit of a full-size lattice, due to the macroscopic spatial coherence of polariton condensates, such connections may extend well beyond a single cell, suggesting a complex network of interactions across the lattice structure.
To maintain the paradigm of pairwise interactions, it is essential to isolate the dyads from each other within the polariton lattice.
This can be achieved by introducing additional potential barriers between condensates whose connections need to be severed.
Figures~\ref{FIG_Scheme1}(h) and~\ref{FIG_Scheme1}(j) illustrate pairs of polariton dyads in OFF and ON states, respectively, with the dyads separated by real potential barriers.
In Figs.~\ref{FIG_Scheme1}(k) and~\ref{FIG_Scheme1}(l), imaginary absorbing barriers are used for the separation.
It is evident that independently of the barrier's nature, the controlling optical beam, in both scenarios, switches both dyads from the OFF state to the ON state, while maintaining the dyads as non-interacting entities, ensuring that their interconnections remain unaffected.

While managing isolation for a single pair of dyads is relatively straightforward, achieving the same for a full lattice of dyads requires a more comprehensive approach.
This involves carefully selecting the appropriate nature and configuration of the potentials to provide sufficient isolation for each dyad on all sides without impeding their interaction with control signals.
Figure~\ref{FIG_ManyDyads} presents the results of numerical simulations for a lattice of dyads, schematically depicted in Figs.~\ref{FIG_Scheme1}(a) and~\ref{FIG_Scheme1}(b), both in the absence [Fig.~\ref{FIG_ManyDyads}(a)] and in the presence [Fig.~\ref{FIG_ManyDyads}(b)] of signal  pulses.
The shape of the potential that ensures the successful operation of each artificial neuron within the lattice is schematically depicted in white in Fig.~\ref{FIG_ManyDyads}(c).
A more detailed analysis of the resulting behavior of artificial neurons under various isolation conditions and interactions with control pulses is provided in Sec.~S3 of the Supplementary Information.
The possible experimental implementation of the proposed system is discussed in Sec.~S6 of the Supplementary Information.

%SSSSSSSSSSSSSSSSS
%SSSSSSSSSSSSSSSSS
%EEEEEEEEEEEEEEEEE
%CCCCCCCCCCCCCCCCC
\subsection*{Network Architecture and Design}

%SSSSSSSSSSSSSSSSS
%SSSSSSSSSSSSSSSSS
%SSSSSSSSSSSSSSSSS
%EEEEEEEEEEEEEEEEE
%CCCCCCCCCCCCCCCCC
\subsubsection*{Hidden layer of the polariton dyad lattice}

Typically, a neuromorphic network comprises in its core an input layer where signals are initially received, processed through a series of intermediate hidden layers.
The output layer then finalizes the processing sequence.
In this paper, we propose a binarized polariton network with a single hidden layer based on a lattice of polariton condensates, see Fig.~\ref{FIG_Scheme1}(a).
For the sake of simplicity, we consider square lattices, characterized by an equal number of nodes $n_{\text{c}}$ along each side, although our results can be generalized to lattices of an arbitrary shape.
Within the lattice, adjacent condensates are randomly paired into dyads, that are mutually isolated, ensuring that the state of each dyad remains independent and it does not affect adjacent pairs.
The dyads in the network function as binary neurons.
They switch between OFF and ON states, corresponding to the output signal values of 0 and 1, respectively, in response to the absence or presence of an additional pumping control signal near each dyad, which acts as an input signal.
Each dyad at the center of the lattice can be subjected to two input signals, while dyads at the edges are influenced by only one signal, cf. \textit{(x)} and other numbered dyads in Fig.~\ref{FIG_Scheme1}(a).
The dyad unaffected by the input signal \textit{(vii)} remains in the OFF state, while the dyads affected by one \textit{(v)} or two \textit{(iv)} input signals switch to the ON state. 
Due to the randomness in pairing condensates within the lattice, some individual condensates may not be part of any dyads, see \textit{(xiii)} in Fig.~\ref{FIG_Scheme1}(a).
In such cases, these  condensates do not contribute to the generation of the output signal.

The random arrangement of dyads within the lattice also results in several options of how the input signal can influence the neurons in the hidden layer.
In the absence of the signal, no neurons are activated, see \textit{(1)} in Fig.~\ref{FIG_Scheme1}(a).
Depending on the number of dyads adjacent to the input signal, it can activate either one~\textit{(2)} or two~\textit{(3)} neurons at once, or none at all~\textit{(4)}.

The square geometry of the lattice dictates the ratios of the numbers of elements and their mutual arrangement within the interconnected sub-lattices.
In particular, to accommodate a ${n_{\text{in}} \times n_{\text{in}}}$ grid of input signals, a ${(n_{\text{in}} + 1) \times (n _{\text{in}} + 1)}$ grid of polariton condensates 
is required.
In such a lattice, the total number of dyads would be~${N_{\text{d}} \le (n_{\text{in}} + 1)^2/2}$. 

In our proposal, the inherent randomness in the pairing of adjacent condensates into dyads within the lattice introduces a significant element of nonlinearity into the network's functioning.
This randomness in dyad formation means that the impact of input signals on the neurons varies, with some dyads being activated by one or two signals, while others remain unaffected.
This randomness leads to a complex network response that cannot be easily mapped or predicted linearly. 
Such variability in dyad responses, triggered by input signals, enhances the computational capabilities of the hidden layer.
The dyads, functioning as binary neurons, switch between OFF and ON states in response to control signals, similar to a classical OR gate, where receiving at least one `1' input signal triggers a `1' output signal.
However, the diverse arrangement of these dyads, some isolated and some interconnected, ensures a dynamic and non-linear processing environment within the network, crucial for tackling intricate computational tasks.
For a more detailed discussion of the nonlinearity induced by the structure of the neural network, see Sec.~S1 of the Supplemental Information.

Using the hidden layer configuration illustrated in Fig.~\ref{FIG_Scheme1}(a,b) as an example, we demonstrate the feasibility of practical implementation of a hidden layer based on a lattice of polariton condensates.
Using a model based on the generalized Gross-Pitaevskii equation for the polariton wave function~$\Psi(t,\mathbf{r})$, as detailed in the Methods section, we conduct a series of numerical experiments to simulate the excitation of a lattice of polariton dyads under control by laser pulses.
The polariton condensates are excited by non-resonant laser pulses with intensity ${P_1 (t,\mathbf{r}) = P_{10} f_1(t,\mathbf{r},w_1,w_{\tau1})}$, arranged in a $5\times5$ lattice configuration as shown in Fig.~\ref{FIG_Scheme1}(a,b).
Here, $P_{10}$ is the intensity magnitude, and $f_1(t,\mathbf{r},w_1,w_{\tau1})$ is a Gaussian function of space and time with parameters of width $w_1$ and  duration~$w_{\tau1}$ of pulses.
The signal pulses with intensity ${P_2 (t,\mathbf{r}) = P_{20} f_2(t,\mathbf{r},w_2,w_{\tau2})}$ are applied to specific nodes in the intervals between the condensate pump pulses.
The meaning of the parameters in ${P_2 (t,\mathbf{r})}$ is analogous to that in~${P_1  (t,\mathbf{r})}$.
The positions of the pump spots are indicated in Fig.~\ref{FIG_ManyDyads}(c).
Shapes of pulses as well as values of the parameters used for simulations are detailed in the Methods section.

For the mutual isolation of dyads, we employ an effective potential, schematically depicted in white in Fig.~\ref{FIG_ManyDyads}(c).
It is important to note, however, that despite its sophisticated shape, there is no need for a unique potential design for each dyad.
The shape of the potential within each dyad is identical for all dyads, and the final lattice configuration is assembled from these uniform building blocks.
Approaches to forming localization potentials for polaritons are discussed in the Discussion section.

In Fig.~\ref{FIG_ManyDyads}, we show time-integrated spatial distribution of the polariton density, $I(\mathbf{r}) = \int |\Psi(t,\mathbf{r})|^2dt$, at certain pump intensities in the absence (a) and in the presence (b) of the the signal pulses.
The parameters of the pulses are chosen such that, in the absence of signal pulses, all artificial neurons remain in the OFF state with a minimum intensity at the center of the dyads, as shown in panel (a).
For the lattice period $d$ taken as 12~$\mu$m, pulses with a duration $w_{\tau1}$ of 5~ps and a width $w_1$ of 2.2~$\mu$m maintain these conditions over a broad range of pump intensities.
In the presence of signal pulses, the proximity of such a pulse to a dyad switches it to the ON state, as shown in panel (b).
When selecting the parameters for the signal pulses, it is essential to ensure that they can effectively influence the polariton dyad during its evolution.
Our simulations show that pulses with a width $w_2=w_1$ and a slightly longer duration $w_{\tau2}$, taken as 8~ps for Figs.~\ref{FIG_ManyDyads}(a,b), are sufficient to achieve this effect.
Additionally, we assume that the signal pulses and the pump pulses arrive synchronously, with their peaks coinciding in time.

Since the pump intensity gradually increases in the pulsed excitation regime, dyads switch to the ON state gradually rather than instantaneously, and they may change their parity during evolution.
Consequently, even in the OFF state, the time-integrated intensity at the center of the dyad can be non-zero.
However, it must always remain lower than the intensity of a dyad in the ON state.
To analyze whether the pump parameters meet this condition, it is useful to introduce the parameter of distinguishability of OFF and ON states, $\Delta \tilde{I}$.
For a polariton lattice with a given configuration, $\Delta \tilde{I}$ represents the difference between the minimum intensity of output signals from dyads expected to be in the ON state, $\text{Min}(I_{\text{ON}})$, and the maximum intensity of signals from dyads expected to be in the OFF state, $\text{Max}(I_{\text{OFF}})$, normalized by the latter, $\Delta \tilde{I} = [\text{Min}(I_{\text{ON}}) - \text{Max}(I_{\text{OFF}})]/\text{Max}(I_{\text{OFF}})$.
A successful choice of pump parameters is characterized by positive values of $\Delta \tilde{I}$.
If $\Delta \tilde{I}$ is zero or negative, the ON and OFF signals become indistinguishable.
The color map in Fig.~\ref{FIG_ManyDyads}(d) outlines the region in the plane of pump pulse intensities $(P_{10},P_{20})$ where ON and OFF states of the artificial neurons are distinguishable, with other parameters held constant.
It is evident that, under the optimal combination of parameters (indicated by the blue line in the panel), the distinguishability $\Delta \tilde{I}$ can reach a value of one, meaning that the difference in intensities is comparable to the intensity of the output signal of the neuron in the OFF state.
Such distinguishability is expected to provide excellent robustness against noise.
It significantly exceeds the corresponding measure in~\cite{NanoLett213715}, which, according to rough estimates, does not exceed 0.2 for a single neuron.

The distribution of polariton density within each dyad represents a time-averaged interference pattern, largely determined by the relative positions of the condensates within the dyad and the localization potential.
We investigated the sensitivity of the output signal of the polariton lattice to deviations in the positions of the condensates from their designated positions.
For this purpose, for several pairs of pump pulse intensities, marked in Fig.~\ref{FIG_ManyDyads}(d) (numbered from $j=1$ to $j=17$), we calculated the output signal intensities of the neurons in the treated example lattice configuration, in the presence of input signals.
In Fig.~\ref{FIG_ManyDyads}(e), we present the normalized output signal intensities for all neurons as a function of~$j$.
The normalization is done with respect to the reference intensity, which for each neuron is the output signal intensity in the absence of the input signal.
For each~$j$, a series of independent numerical experiments was conducted, where the positions of the pump spots were randomly deviated from their designated positions within the range from $-\delta$ to $+\delta$.
Each point on the individual panels in Fig.~\ref{FIG_ManyDyads}(e) corresponds to a separate numerical experiment, with different colors representing different magnitudes of deviation~$\delta$.
It can be observed that random deviations within approximately $\pm 0.5 \, \mu \text{m}$ do not undermine the distinguishability of the neuron states.

%SSSSSSSSSSSSSSSSS
%SSSSSSSSSSSSSSSSS
%SSSSSSSSSSSSSSSSS
%EEEEEEEEEEEEEEEEE
%CCCCCCCCCCCCCCCCC
\subsubsection*{Preparation of an input layer  \label{SSecPrepInpLay}}

As previously mentioned, the input signals in our system are generated by non-resonant optical beams that modify the potential landscape near the dyads.
These beams excite incoherent exciton reservoir spots, that influence both the real and imaginary parts of the effective potential.
The beams are generated by a set of identical emitters, e.~g., using a spatial light modulator~\cite{Optica8106,Nanolett203506}, and are systematically arranged according to a  pattern derived from the initial signal, ensuring that the input beams align accurately with the required configuration for the neural network's processing.

The process for preparing the input signal pattern is outlined in a sequence of steps as follows, demonstrated using the example of classifying handwritten digits from the MNIST dataset, see Fig.~\ref{FIG_Scheme2}.
The initial image is transformed into a matrix with dimensions $n_0 \times n_0$, mirroring the image's size.
Each element of the matrix corresponds to the grayscale level of its respective image pixel.
In the MNIST dataset, $n_0 = 28$, and the total number of pixels is~${N_{0} = n_0^2 = 784}$.
Next, the matrix is binarized by rounding up its values.
At this step, the size of the matrix remains unchanged.

\begin{figure}[tb!]
\begin{center}
\includegraphics[width=0.9\linewidth]{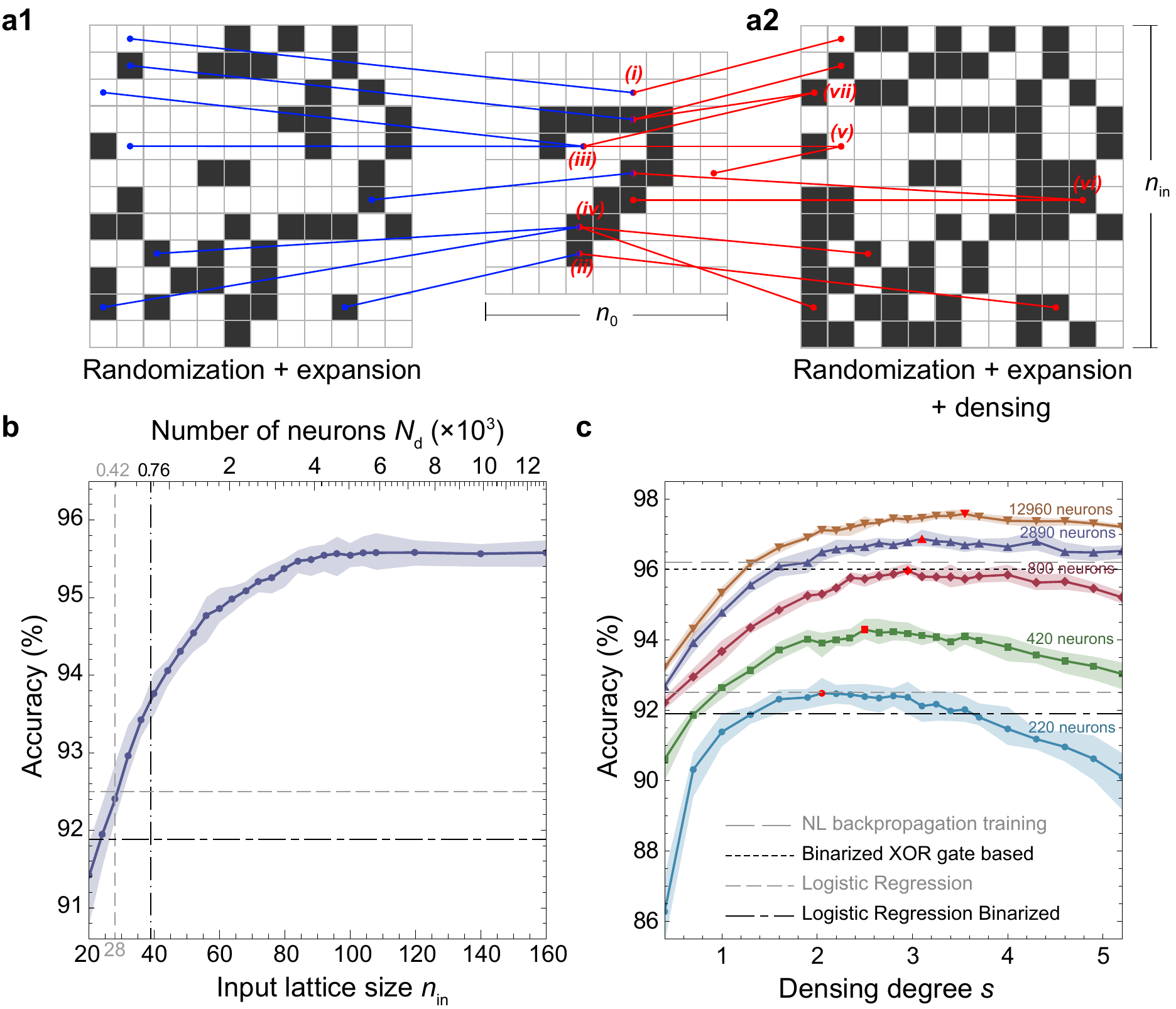}
\end{center}
\caption{
{
Evaluation of the MNIST handwritten digit recognition by the polariton neuromorphic network.
(a1,a2)~Schematic depicting the conversion of a binarized initial signal lattice of size $n_{0} \times n_{0}$ into an input signal lattice of size $n_{\text{in}} \times n_{\text{in}}$, utilizing randomization and expansion, both without (a1) and with (a2) signal densing.
(b)~The recognition accuracy in dependence on the size of the input signal lattice $n_{\text{in}}$ (lower scale) or the number of neurons (dyads) in the hidden layer $N_{\text{d}}$ (upper scale).
Each data point represents an average from ten numerical experiments, each utilizing different randomization masks.
The shaded area reflects the variation in accuracy across these numerical experiments.    
Vertical lines, serving as guides for the eye, indicate the conditions, where the size of the polariton lattice matches that of the initial images (dashed), and when the number of neurons equals the number of pixels in the initial image (dash-dotted).
(c)~The recognition accuracy in dependence of the densing degree $s$ of the input signal for square polariton lattice systems of different size $d_{\text{in}}$ with different numbers of neurons in the interaction layer: $d_{\text{in}} = 20$ with 220 neurons (blue),  $d_{\text{in}} = 28$ with 420 neurons (green),  $d_{\text{in}} = 39$ with 800 neurons (red), $d_{\text{in}} = 75$ with 2890 neurons (violet), and  $d_{\text{in}} = 160$ with 12960 neurons (brown).
Red markers indicate the maxima of the dependencies.
Horizontal dashed lines indicate the accuracy levels for alternative established classification approaches (from bottom to top): the linear software classification of the grayscale~(92.5\%) and binarized (91.9\%) MNIST data set, the binarized polariton network based on a layer of XOR gates~\cite{NanoLett213715} and the nonlinear polariton network with the software backpropagation training~\cite{PhysRevApplied18024028}.
}
\label{FIG_ResultsMNIST}}
\end{figure}

On the next step, the pattern preparation takes place, which involves several operations simultaneously, see Fig.~\ref{FIG_ResultsMNIST}(a1).
The first operation is randomization.
The elements of the binarized matrix are transferred to the pattern matrix not sequentially, but in a random manner, using a randomization mask that is consistent for all recognized images.
This technique allows for a more uniform distribution of the signal across the entire layer, engaging both central and peripheral neurons to the same extent.
Such an approach ensures that the neural network utilizes its entire structure more effectively, enhancing overall performance and accuracy.

The second operation involves expansion, which entails increasing the number of neurons engaged in processing.
This step enhances the network's capacity to handle and interpret complex data by involving a larger array of neurons, thereby improving its computational power and efficiency.
In the geometry of the square lattice under consideration, the number of neurons in the hidden layer should increase superlinearly with the number of elements in the input signal lattice, $N_{\text{in}} = n_{\text{in}}^2$, as $N_{\text{d}} \le (N_{\text{in}} + 2 \sqrt{N_{\text{in}}} +1)/2$.
To increase the lattice size from $n_0$ to $n_{\text{in}}$, elements randomly selected from the binarized lattice are repeatedly inserted into the input pattern lattice, see \textit{(iii)} and \textit{(iv)} in Fig.~\ref{FIG_ResultsMNIST}(a1).
One should note that both randomization and expansion are uniformly applied to all initial images, ensuring that each element of the binarized matrix is allocated to specific, unchanging positions within the input pattern lattice, consistent across different images. 
This approach to element placement standardizes the processing framework for each image.

%FFFFFFFFFFFFFFFFF
%IIIIIIIIIIIIIIIII
%GGGGGGGGGGGGGGGGG

%SSSSSSSSSSSSSSSSS
%SSSSSSSSSSSSSSSSS
%EEEEEEEEEEEEEEEEE
%CCCCCCCCCCCCCCCCC
\subsection*{Network operation and accuracy evaluation}

%SSSSSSSSSSSSSSSSS
%SSSSSSSSSSSSSSSSS
%SSSSSSSSSSSSSSSSS
%EEEEEEEEEEEEEEEEE
%CCCCCCCCCCCCCCCCC
\subsubsection*{MNIST dataset analysis}

The pattern resulting from the previous step serves as the guide for the spatial light modulator, which then generates the input signal for the hidden layer of the binarized polariton network.
In the hidden layer, the key part of computational processes takes place, culminating in the generation of the output signal, see Fig.~\ref{FIG_Scheme2}.
It is important to emphasize that due to the multiple interaction possibilities between neurons of the input layer and the neurons of the hidden layer, each element can potentially activate a different number of neurons in the hidden layer, depending on its position within the signal lattice.
This highlights the nonlinear nature of these interactions.
The output signal is composed of an array of binary signals, either 0 or 1, which encode the OFF and ON states of polariton dyads in the hidden layer. These states are manifested as the presence or absence of photoluminescence from the center of the dyads.
The output signal then can be processed using a conventional linear classifier, which can be implemented either electronically or through purely optical means~\cite{NanoLett213715,PhysRevApplied16024045}.

To evaluate the effectiveness of the proposed polariton neuromorphic network architecture, we simulated its operation using numerical calculations.
The assessment was conducted on the MNIST dataset~\cite{lecun1998gradient,lecun2010mnist}, comprising a set of $60000$ samples of handwritten digits for training the network and a set of $10000$ samples for testing purposes.
Given that the MNIST dataset serves as a standard benchmark for the goals of image classification, its use in our study enables a rigorous comparison of our findings with existing results in polariton-based neuromorphic network research~\cite{Nanolett203506,PhysRevApplied18024028,NanoLett213715,PhysRevApplied16024045}.
It is noteworthy that the binary nature of both the input and output signals, as well as the functioning of the neurons within the hidden layer, facilitated a more efficient utilization of computational resources. 
This efficiency, in turn, permitted the deployment of the entire spectrum of training and testing images, eliminating the need for any artificial reduction in image resolution.
At the linear classifier stage of the neural network, the linear regression algorithm 
was employed for efficient data classification.

It is commonly observed in the neuromorphic network research that the increase of the number of neurons in the hidden layer can potentially improve the classification accuracy.
This is corroborated by our findings, as illustrated in Fig.~\ref{FIG_ResultsMNIST}(b), where we demonstrate a monotonic increase in the accuracy of classifying MNIST dataset images with the increase in the size of the input signal lattice~$n_{\text{in}}$, which implies an increase in the number of neurons~$N_{\text{d}}$ in the hidden layer.
In the figure, each data point corresponds to the accuracy of the image recognition at a given number of neurons, averaged over ten numerical experiments, carried out for different randomisation masks used at the input layer preparation step.
The light blue shaded area indicates the range of variation of accuracy obtained in the corresponding numerical experiment series.
It can be observed that randomization has a finite effect on accuracy, altering it within a margin of less than one percent.
This dependence exhibits a saturating character, converging towards about 95.4\% accuracy with an increasing number of neurons.
Vertical lines indicate the observations, corresponding to the size of the input layer equal to the size of the initial image, ${n_{\text{in}} = n_0 = 28}$, with the number of neurons estimated as about 420 (gray dashed), and to the size of the input layer ${n_{\text{in}} = 39}$, with the number of neurons close to the number of pixels in the initial image, $N_{\text{d}}=760 < n_0^2$ (black dot-dashed). 
It can be seen that in the second case, even with a slightly smaller number of neurons involved, the classification accuracy of the polariton neuromorphic network, averaging over 93.3\%, exceeds the accuracy of the software linear classification of both grayscale (92.5\%) and binarized (91.9\%) images, see also Sec.~S2 in the Supplementary Information.

In addition, reducing the input signal lattice dimension to~20, and consequently the number of neurons in the hidden layer to approximately~220, still ensures an accuracy above 90\%.

%SSSSSSSSSSSSSSSSS
%SSSSSSSSSSSSSSSSS
%SSSSSSSSSSSSSSSSS
%EEEEEEEEEEEEEEEEE
%CCCCCCCCCCCCCCCCC
\subsubsection*{Input signal densing}

In the procedure previously described above, it was assumed that the value of each element in the input layer (0 or 1) is determined by the value of only one randomly selected element from the binarized initial image.
Herewith, the average filling of the binarized initial matrix as well as of the input matrix (number of nonzero elements relative to the total number of elements) is about 0.19.
Thus, on average, more than 80\% of the input neurons do not trigger the activation of neurons in the hidden layer.
Meanwhile, the nonlinearity of the interaction is manifested only in the case of activated neurons.

To further increase accuracy, we suggest to supplement the preparation of the input layer with an additional operation, termed by us \textit{input signal densing}, which allows to increase the filling of the input layer and thereby enhance the nonlinearity.
The operation involves using more than one element of the binarized matrix layer for determining the value of an element in the input layer.
In this scenario, when projecting the binarized image onto the input signal layer, the selection process is randomized not only for the element being transferred from the first matrix but also for the element in the second matrix where the transfer occurs.

In the approach not involving the image densing, all elements can be filled consecutively.
Once a particular element in the input layer is assigned a value, this value remains unchanged, and the element no longer participates in the further process of filling the input layer.
Each such element contains only one incoming connection from the projected binarized layer, see blue lines in Fig.~\ref{FIG_ResultsMNIST}(a1).
In contrast, in the case of image densing, the value of a filled element can be modified during the subsequent process.
This approach is illustrated in Fig.~\ref{FIG_ResultsMNIST}(a2).
If two elements of the binarized image matrix are used for determining an element of the input matrix, in this case, the principle of assigning values is similar to the operation of a logical OR gate.
If both elements are 0, the assigned element in the input layer also takes a value~0, see \textit{(v)} in Fig.~\ref{FIG_ResultsMNIST}(a2).
If either one of the elements \textit{(vi)} or both of them \textit{(vii)} are 1, then the assigned element is also~1.
This can be easily generalized to the case of three and more elements.

In the proposed approach, the preparation of the input layer involves partial overlapping of the initial image with itself.
To describe such overlapping, we introduce the parameter of densing degree~$s$, which can take both integer and fractional values.
For instance, a densing degree of 2 means that each element in the input layer is on average determined by two elements of the initial binarized layer.
Similarly, with ${s=1.5}$, approximately half of the input layer elements are on average determined by one element of the initial layer, while the other half by two elements.
Figure~\ref{FIG_ResultsMNIST}(c) illustrates how the classification accuracy of the polaritonic network, with different numbers of neurons in the hidden layer, changes with the densing degree $s$ of the input layer.
The figure reveals pronounced peaks in the accuracy dependencies.
This phenomenon can be explained as follows.
States both in the input space and in the feature space are represented by binary vectors consisting of various combinations of `0's and~`1's.
Increasing the densing degree $s$ implies a higher fraction of `1's in the input space, which in turn activates more neurons in the hidden layer.
This activation facilitates structural nonlinearity through the network's architecture, thereby enhancing the distinct features of elements belonging to specific data classes (digits in our case).
Consequently, this also increases the number of `1's in the feature space.

However, beyond a certain point, further increasing the densing degree $s$ results in diminishing returns.
Additional `1' in the feature space no longer contribute to highlighting distinctive features.
Instead, they may obscure these features by overwhelming other state elements or adding redundant information.
In the extreme case where $s$ approaches infinity, every input state would be mapped to a feature space vector consisting entirely of `1's, rendering state recognition impossible.
For a given number of involved neurons, the peak accuracy is achieved when the number of `0's and `1's is comparable with each other.
The variability in the peak's position and the rate at which classification accuracy changes with different numbers of neurons in the hidden layer can be attributed to the structural nonlinearity introduced by the random pairing of input signals with neurons.
This nonlinearity varies with each configuration, significantly impacting how effectively the network processes and utilizes the increased density of `1's to distinguish between classes.
Consequently, the network's ability to manage this increased input density depends on the specific configuration and number of neurons.

The key finding of our study is that with approximately 800 neurons (corresponding to ${n_{\text{in}}=39}$) and a densing degree of about~3, the accuracy reaches the previously reported benchmark of 96\%~\cite{NanoLett213715,PhysRevApplied16024045} (see red marker on a red curve in Fig.~\ref{FIG_ResultsMNIST}(c)), and surpasses it with the further increase of the number of neurons.
Our predicted maximum average accuracy is 97.5\%, obtained for the network with the input matrix size of ${n_{\text{in}} = 160}$, corresponding to the number of neurons less than~$1.3\times 10^4$, and the densing degree ${s=3.5}$, see the red marker at the brown curve in Fig.~\ref{FIG_ResultsMNIST}(c).
The accuracy can be further increased by at least 0.1\% through the appropriate choice of a randomization mask.
It is evident that the dependencies exhibit a pronounced maximum, and the value of $s$, at which this maximum occurs, increases with the growing number of neurons~$N_{\text{d}}$.
Moreover, the fewer the number of neurons, the steeper the dependencies become.
It can also be observed that the impact of the randomization mask notably  decreases with an increase of the number of neurons.

%FFFFFFFFFFFFFFFFF
%IIIIIIIIIIIIIIIII
%GGGGGGGGGGGGGGGGG
\begin{figure}[tb!]
\begin{center}
\includegraphics[width=0.9\linewidth]{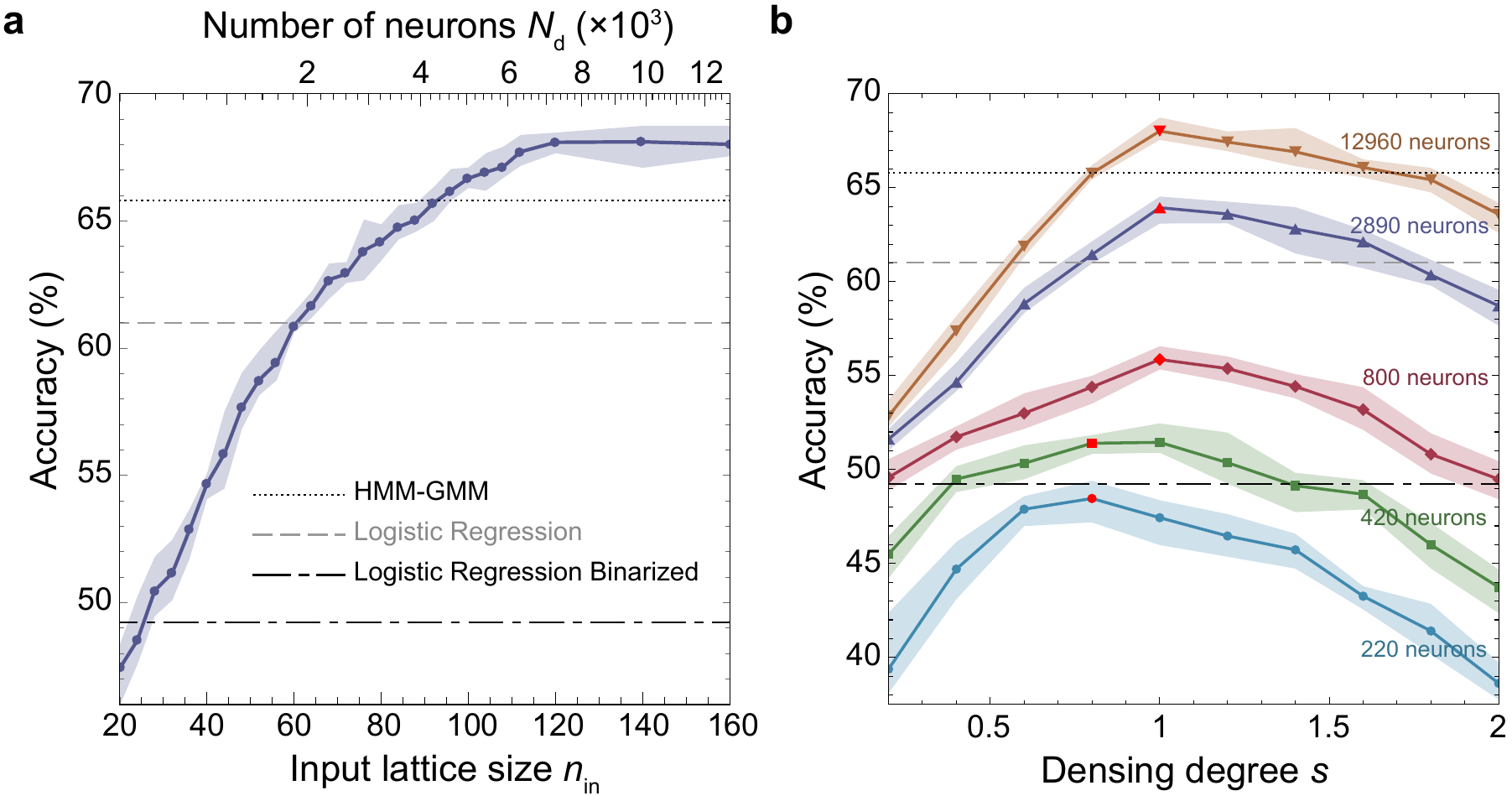}
\end{center}
\caption{
{
Evaluation of the Speech Commands~\cite{warden2018speechcommandsdatasetlimitedvocabulary,speechcommandsv2} recognition (ten commands) by the polariton neuromorphic network.
(a)~The recognition accuracy in dependence on the size of the input signal lattice $n_{\text{in}}$ (lower scale) or the number of neurons (dyads) in the hidden layer $N_{\text{d}}$ (upper scale).
(c)~The recognition accuracy in dependence of the overlap degree $s$ of the input signal for square polariton lattice systems of different size $d_{\text{in}}$ with different numbers of neurons in the interaction layer: $d_{\text{in}} = 20$ with 220 neurons (blue),  $d_{\text{in}} = 28$ with 420 neurons (green),  $d_{\text{in}} = 39$ with 800 neurons (red), and  $d_{\text{in}} = 160$ with 12960 neurons (brown).
Red markers indicate the maxima of the dependencies.
Horizontal dashed lines indicate the accuracy levels for alternative  classification approaches (from bottom to top): the linear software classification of the binarized~(49.2\%) and non-binarized (61\%) MFCC feature matrices, and the HMM-GMM-based classification of the MFCC feature matrices (65.8\%).
}
\label{FIG_ResultsSound}}
\end{figure}

{
%SSSSSSSSSSSSSSSSS
%SSSSSSSSSSSSSSSSS
%SSSSSSSSSSSSSSSSS
%EEEEEEEEEEEEEEEEE
%CCCCCCCCCCCCCCCCC
\subsubsection*{Speech Commands dataset analysis}

To further evaluate the adaptability of our proposed neuromorphic network and its ability to generalize across different types of data beyond the MNIST dataset, we tested it using the Speech Commands dataset~\cite{warden2018speechcommandsdatasetlimitedvocabulary,speechcommandsv2}.
This dataset fundamentally differs from MNIST in structure, consisting of one-second audio clips containing spoken words rather than static images.
The Speech Commands dataset is designed for voice recognition tasks, featuring audio recordings of spoken commands.
Focusing on ten classes --- digits from zero to nine --- we converted the audio files into flattened one-dimensional arrays of Mel-frequency cepstral coefficients (MFCCs), which are common features used in speech recognition~\cite{IEEE1163420,SAHIDULLAH2012543}. 
Each MFCC feature matrix comprises 12 coefficients that capture the primary audio characteristics.
For binary processing compatibility, these matrices were binarized so that non-positive elements were set to `0' and positive elements to~`1'.
For detailed information on the preprocessing steps and the composition of the MFCC feature matrix, refer to the `Methods' section and Supplementary Section~5.

Expanding our evaluation to the Speech Commands dataset, our results are detailed in Fig.~\ref{FIG_ResultsSound}, showcasing how the network adapts to audio recognition tasks.
As anticipated, the classification accuracy for spoken digits increases with the number of neurons, peaking at about 68\%, with results presented in Fig.~\ref{FIG_ResultsSound}(a).
Similar to our findings with image-based datasets, the performance of our neuromorphic network excels beyond that of traditional linear classifiers both in the binarized (by about~19\%) and non-binarized (by about~7\%) MFCC feature arrays.

In examining the performance of our proposed neuromorphic network on the Speech Commands dataset, it's beneficial to compare it against a well-established benchmark in speech recognition: the Hidden Markov Model with Gaussian Mixture Model (HMM-GMM)~\cite{IEEETrSP134406,SIG0042008}.
This approach utilizes Hidden Markov Models to sequence audio data, where each state is characterized by a mixture of Gaussian distributions.
Traditionally used for modeling audio sequences in speech recognition tasks, HMM-GMM operates on input data without undergoing binarization, providing a robust baseline with a reported accuracy of 40.9\% across the full Speech Commands dataset~\cite{rai2023keywordspottingdetecting} and 65.8\% for the ten-class subset.
Impressively, our binary neuromorphic network surpasses this result, achieving even greater accuracy on binarized data.

A notable observation from our study pertains to the application of the input signal densing technique.
As depicted in Fig.~\ref{FIG_ResultsSound}(b), unlike with the MNIST dataset, adjusting the densing degree $s$ for the Speech Commands does not yield improvements in classification accuracy, regardless of the neuron count.
The ineffectiveness of densing in this context can be attributed to the intrinsic data structure of Speech Commands, where the average filling of the initial binarized matrix is approximately 0.51, closely balancing the `0' and `1' signals.
This balance is critical because the linear classifier used for post-processing in the network treats imbalances towards more 0's or more 1's with similar effectiveness.

For datasets characterized by a denser filling, where the majority of the input vector elements are active, achieving a balanced input suitable for optimal classifier performance can often be managed at the binarization stage by adjusting the threshold.
Here, the threshold is the value at which pixels in the initial data vector are converted to binary values: pixels with intensities above this threshold are set to `1', and those below or equal to the threshold are set to `0'.
In the case of MNIST, where the average natural filling is significantly lower (around 0.19), achieving such balance using threshold adjustments alone is not feasible, thus highlighting the utility of the input signal densing technique.

It is somewhat presumptuous to specify an exact optimal filling degree for the input vector because in fact the parity should be achieved in the output signal entering the linear classifier, following its traversal through the hidden layer.
Nevertheless, the balance of 0's and 1's in the input signal can serve as a rough guideline for approximating this parity.

Further details on the applicability of both threshold management and the input signal densing technique, particularly with the MNIST datasedt, are discussed in Supplementary Section~S6. 
Supplementary Section~S4 provides accuracy estimates for the Fashion-MNIST dataset, where a zero threshold in binarization (assigning a `1' to any non-zero pixel) naturally leads to a filling that approaches parity.
These insights underscore the boundaries and effectiveness of the input signal densing technique across different datasets.
}

%SSSSSSSSSSSSSSSSS
%EEEEEEEEEEEEEEEEE
%CCCCCCCCCCCCCCCCC
\section*{Discussion}

Polaritonic neural networks, both previously proposed and presented in this work, are constructed based on a single hidden neuronal layer.
In this context, the nature of neuronal connections plays a crucial role in determining the accuracy of such a network.
In Ref.~\cite{NanoLett213715}, optical XOR gates function as neurons in the hidden layer.
Input to each gate consists of a pair of optical pulses encoding two random pixels from a binarized initial image, and the output is the result of nonlinear interaction of these pulses.
Thus, in this architecture of the neural network, nonlinearity is achieved at the level of interaction of the input layer neurons, while the triggering of neurons in the hidden layer merely reflects the outcome of these interactions.

In the architecture without input signal densing, proposed by us, input neurons do not interact with each other.
Nonlinearity is achieved at the level of interaction between the input neurons and the neurons in the hidden layer.
Thus, the neurons in the hidden layer not only transmit the result of the interaction but are themselves the subjects of this interaction.

Our proposed neural network architecture is highly advantageous as it allows for the integration of both approaches to facilitate nonlinear interactions among neurons of different layers.
The integration was realized through the introduction of an input signal densing procedure, leading to record-breaking level of predicted accuracy in polaritonic neural networks, substantially surpassing those of previously proposed architectures.

The path towards further enhancing the accuracy of our system remains open.
Due to the flexible geometry of the lattice and the genuinely nonlinear nature of the interactions among the lattice-forming polariton condensates, there is potential to increase accuracy by involving a larger number of neurons from the hidden layer in contact with input signals.
Additionally, a modification of the nature of neuronal interactions, such as by replacing the OR operation with a XOR operation in the hidden layer, could also contribute to the improvement of accuracy.

To address the challenge of scaling polariton neural networks beyond current limitations, we propose several strategies.
These strategies, detailed in Sec. S4 of the Supplementary Information, include tiling multiple spatial light modulators (SLMs), utilizing optical waveguides, implementing advanced micro-optics, and employing dynamic reconfiguration techniques.
By using multiple SLMs placed adjacent to each other, one can create a larger composite grid that significantly increases the addressable area for the pump beams.
This modular approach allows for the expansion of the network by simply adding more SLMs, with careful alignment and synchronization ensuring seamless operation across the entire grid.
Optical waveguides can be integrated into the sample to direct light precisely to designated spots, overcoming the spatial limitations of free-space optics.
These waveguides can be fabricated using advanced lithography techniques to create efficient light paths with minimal loss, enabling the creation of larger networks without increasing the physical footprint.

Incorporating advanced micro-optics, such as microlens arrays, allows for precise focusing and directing of pump beams.
This enhances the density of pump spots within the available area, enabling more neurons to be addressed simultaneously.
Techniques like diffractive optical elements can further optimize the beam distribution and intensity profile.
Dynamic reconfiguration techniques enable a single SLM to sequentially address different regions of the sample at high speeds or to reproduce different configurations of the neural network on the same region.
This time-multiplexing approach effectively increases the number of controllable neurons without increasing the physical size of the SLM or the sample, by dynamically reconfiguring the pump spots or the neural network fragments.
Each of these approaches is designed to overcome spatial constraints and enhance the efficiency of pump beam delivery, thereby enabling the creation of larger and more complex polariton neural networks.
For more detailed descriptions and technical aspects of these strategies, refer to the Supplementary Information.

The main advantage of the polaritonic part of the neural network is its unprecedented speed, with processing times of only a few tens of picoseconds required for the generation and evolution of polariton condensates in response to laser pump pulses.
Delays caused by electronic components are inevitable for neural networks of any nature during data loading and result reading stages.
However, all other stages in a polaritonic network can be realized without electronic components.

In each specific experiment, the input signal transformation masks, once set initially, remain unchanged throughout the experiment.
This means that while the input signals change, the optical paths of individual signals remain constant.
Among the operations mentioned, randomization and expansion are linear operations that can be described as matrix-vector multiplications.
These operations can be optically implemented, as mentioned in~\cite{Nanolett203506,PhysRevApplied18024028} and detailed in~\cite{SciRep812324,OptLett455752}.
An input light field representing the vector of input signals passes through an optical element that applies a spatially varying phase or amplitude modulation, effectively performing a matrix-vector multiplication at the speed of light.

The signal densing operation is more complex, but solutions exist for its optical implementation as well.
Our studies on the dependence of neuron performance on signal pulse intensity show that, with other parameters fixed, the intensity of signals can vary within several folds while ensuring correct neuron switching and not compromising the isolation of dyads.
In this mode, signal densing can be achieved by simply combining multiple pulses at a single point on the sample, resulting in a proportional increase in the intensity of the resulting pulse.
Such a combination can be realized using optical components like beam splitters and combiners.

Another approach involves temporal multiplexing, where the steps of the signal densing process are spread out over time.
This technique is similar to the dynamic reconfiguration strategy discussed for network scaling, where different steps of the process occur sequentially in time but within the same spatial region.
By carefully timing these steps, we can ensure that the combined effect of the pulses is achieved without significant delay. Even with an increase in the resulting operation time to hundreds of picoseconds, this approach still maintains a substantial speed advantage over electronic neural networks.

Transitioning to the aspects of practical realization of the proposed architecture, it becomes essential to consider the possible experimental realization of polariton condensate lattices designed above. 
Spatially-distributed systems comprising chains and lattices of interconnected quantum entities have gained recognition as platforms for information storage, transmission and processing, as well as simulators of complex phenomena~\cite{PhysRevA89033828,PhysRevB95235301,Science333996,NatPhys8267,CompRenPhys17934}.
Polariton lattices, a recent breakthrough in spatially-distributed quantum systems, are notable for their exceptional spatial coherence~\cite{PhysRevLett124207402,NJPhys19125008,Optica8106,NatMater161120}.
This coherence significantly exceeds that of individual condensates and it facilitates phase locking of nodes across the entire lattice.
This widespread phase synchronization could potentially enhance the nonlinearity of interactions essential for neuromorphic network operation, see, e.~g.,~\cite{Nanolett203506}.
However, this scenario leads to the loss of a key advantage emphasized in our work: the feasibility of neuron response binarization would be significantly compromised, impacting computational speed and resource efficiency.
To address this issue while retaining the advantage of polariton lattices, which is the optical controllability of connections between condensates, we propose ensuring pairwise interactions of condensates within the lattice by selectively severing non-contributing links.
For this purpose, a variety of approaches exists.

The first possible approach involves optical induction of potential barriers for ballistic polaritons within a planar microcavity, similar to the excitation of the polariton lattice itself as well as the input signals.
For this purpose, the non-resonant excitation of the exciton reservoir can be used~\cite{PhysRevB85235102}. %~\cite{PhysRevB85235102,ApplPhysLett101261116}
We propose to employ it for both the polariton lattice and the input signals generation. 
However, it should be noted that in this scenario, the pump will contribute not only to the separation of the condensates but also to the changes in their occupation numbers.
This factor can be mitigated by using a reservoir of dark excitons as a barrier~\cite{PhysRevLett122047403,PhysRevB108165411}.
Dark excitons do not participate in optical interactions and do not directly influence the population of polariton condensates.
Meanwhile, the strong repulsive nature of polariton-exciton interactions is equally characteristic of bright and dark excitons.
A recent paper~\cite{PhysRevB108165411} demonstrates the feasibility of excitation of a dark exciton reservoir through the two-photon absorption.
Given that this approach results in record-long exciton lifetimes, over 20~ns, it suggests that such a reservoir would not contribute to replenishing the polariton condensate.

An alternative approach, described in Ref.~\cite{AdvQT31900065}, consists in the separation of condensates within the lattice through the creation of spatially varying dissipation profiles by controlling the decay rates of polaritons at different lattice sites.
Among the experimental methods mentioned in~\cite{AdvQT31900065}, one is proton implanting into quantum wells, which enables independent control of exciton and cavity photon energies, influencing polariton decay rates~\cite{RepProgrPhys80016503}.
Additionally, controlled stress applied to the substrate can create spatial traps, affecting the coupling of exciton and photon states, thus varying polariton lifetimes~\cite{Science3161007}.
For dynamic dissipation control, electrical carrier injection can be used, causing localized losses through the absorption by excited states~\cite{NatMat151061}. One can also exploit the biexciton formation regime to alter polariton interactions~\cite{PhysRevB62R7763}.
As numerical simulations illustrated in Fig.~\ref{FIG_Scheme1}(h--l) show, both potential landscape profiling and dissipation control are comparably effective tools of the condensate-condensate coupling control.
The choice of the appropriate method then would depend primarily on the experimental capabilities available and the specific goals set for the experiment.

The previously described approaches for the pairwise coupling of condensates in a lattice primarily involved operations within a planar microcavity.
However, traditional methods of structuring cavities, such as deep etching techniques, offer alternative avenues for exploration, allowing, e.~g., for the creation of clusters~\cite{PhysRevLett108126403} %~\cite{PhysRevLett108126403,PhysRevX5011034}
and chains~\cite{PhysRevLett116066402} of micropillars.
Structures, crafted through precise etching processes, may form a distinct spatial arrangement within the microcavity plane. 
Another approach for clustering polariton condensates within a lattice, detailed in Ref.~\cite{PhysRevResearch3023167}, involves creating controllable Josephson junctions of the condensates.
This is achieved through nanostructuring of cavity mirror surfaces via direct laser writing~\cite{EPL13054001}, creating local potential minima, and dynamically tuning the potential landscape using a thermo-responsive polymer affected by a heating laser to vary the optical medium's refractive index.
This method allows for controlled polariton tunneling between condensates, facilitated by finite height potential barriers, leading to the formation of condensates in a thermo-optically adjustable potential landscape.

Our investigation into binary polariton neural networks has demonstrated their fundamental operational principles and potential for energy-efficient computing, suitable for certain applications where high speed and low power consumption are critical. 
Energy consumption and efficiency analysis of the proposed system are discussed in Sec.~S5 of the Supplementary Information.
While binary networks inherently trade off precision for efficiency, our results with a basic logistic regression model on the MNIST dataset have shown promising accuracy levels.
Moreover, the proposed network architecture, while simple, suggests several avenues for enhancement that could address tasks requiring higher precision.

We propose that extending the network's complexity through additional hidden layers could amplify its computational power.
To address the challenge of integrating multiple hidden layers in a polaritonic neural network, the output signals from one hidden layer can be utilized as a template for generating input signals for the next layer.
This approach may involve electronic components for signal regeneration or amplification, which, while enhancing accuracy, compromises the system's computations speed advantage.

Alternatively, integrated optical waveguides for polaritons can link neurons across hidden layers.
These waveguides utilize the bistability effect, where a polariton condensate in a low-intensity state is switched to a high-intensity state by an input trigger~\cite{PhysRevLett101016402}.
The hidden layer photoluminescence signals can serve as these triggers, effectively generating signals in neurons at the opposite ends of the waveguides in subsequent layers.
This approach maintains the speed advantage by avoiding electronic mediation and capitalizes on the intrinsic properties of polaritons for rapid signal transmission.
This approach, although requiring adjustments to the hidden layer geometry, offers a feasible and efficient solution for multi-layer integration in polaritonic neural networks.

Adjusting the network's geometric layout from a square to a hexagonal lattice could further improve its performance by increasing the number of neurons each input signal can potentially activate, thus enhancing the structural nonlinearity of the network.
Moreover, replacing the OR gate response with other types of logical operations, such as XOR gates, could add another layer of nonlinearity.
An XOR gate could toggle the state of each neuron in response to paired inputs without changing the overall parity, introducing a dynamic component to the neuron's response.

We also propose the  idea of accommodating continuous input signals alongside binary outputs.
This approach would utilize the varying intensities of signals, such as those from grayscale images, to modulate input signals, enabling a more nuanced response based on pixel brightness.
By doing so, we could significantly enrich the input feature space without compromising the inherent advantages of binary systems, such as low memory usage and high processing speeds.

Finally, exploring controlled interactions between neurons --- what is often considered a parasitic effect of crosstalk --- could be harnessed beneficialthily.
By fine-tuning the isolation and interaction among neuron pairs, we could potentially enhance the network’s robustness and precision.

%SSSSSSSSSSSSSSSSS
%EEEEEEEEEEEEEEEEE
%CCCCCCCCCCCCCCCCC
%\section{Conclusions}
In summary, we have developed a neuromorphic network architecture leveraging lattices of exciton polariton condensates.
The design takes advantage of a binary framework, where each neuron, facilitated by the spatial coherence of pairwise coupled condensates, performs binary operations.
This coherence ensures efficient network-wide communication, with the binary neuron switching driven by nonlinear repulsion through the excitonic component of polaritons.
The binary nature of a network offers computational efficiency and scalability advantages, setting this system apart from conventional continuous weight models and sequential binary neuromorphic systems.

The network's effectiveness was demonstrated using the MNIST dataset for handwritten digit recognition. Our network has not only shown competitive performance against existing systems, but also surpassed them when taking advantage of the original signal densing technique.
The developed approach allowed the network to  achieve a remarkable 97.5\% classification accuracy, theoretically.
{Further validation was conducted on the Speech Commands dataset, which contains diverse and complex one-second audio clips of spoken words.
This additional testing phase highlighted the adaptability and robustness of our architecture in processing intricate audio data and handling a variety of speech recognition tasks.}

By employing a binary operational framework and exploring various lattice structuring techniques, this study opens new pathways for developing efficient, scalable, and high-speed neuromorphic systems.
We are confident that polaritonic systems have high potentiality as creating powerful tools for complex pattern recognition and data processing tasks.

%SSSSSSSSSSSSSSSSS
%EEEEEEEEEEEEEEEEE
%CCCCCCCCCCCCCCCCC
\section*{Methods}

%SSSSSSSSSSSSSSSSS
%SSSSSSSSSSSSSSSSS
%EEEEEEEEEEEEEEEEE
%CCCCCCCCCCCCCCCCC
\subsection*{Numerical simulation of polariton dyads}

For simulating macroscopic coherent states of polariton dyads, we use the model proposed in Ref.~\cite{PhysRevLett124207402}.
We solve the generalized Gross-Pitaevskii equation for the polariton wave function $\Psi(t,\mathbf{r})$:
\begin{equation}
\label{EqGPE}
i\hbar \partial _t \Psi (t,\mathbf{r}) = \left[-\frac{\hbar ^2 }{2 m^*} \nabla^2 + U(t,\mathbf{r}) - \frac{i \hbar \gamma}{2} \right] \Psi (t,\mathbf{r}),
\end{equation}
where $m^*$ is the effective polariton mass, $\gamma$ is the polariton decay rate.
$U(t,\mathbf{r})$ is an effective potential for polaritons that can be written as 
\begin{equation}
\label{EqPotential}
U(t,\mathbf{r}) = \alpha |\Psi (\mathbf{r})|^2 + U_{\text{d}}(t,\mathbf{r}) + U_{\text{in}}(t,\mathbf{r}) + V(\mathbf{r}).
\end{equation}
The first term in the right hand side of Eq.~\eqref{EqPotential} is responsible for polariton-polariton interactions with the interaction constant~$\alpha$.
The second term given as
\begin{equation}
\label{EqPotentiald} U_{\text{d}}(t,\mathbf{r}) = \left(\frac{g_{1} + i R_1 }{2(\gamma_{\text{R}} + R_1 |\Psi (t,\mathbf{r})|^2)} + G_1 \right) P_1 (t,\mathbf{r})
\end{equation}
characterizes the complex effective potential arising from the excitation of polariton condensates in a dyad through non-resonant pulsed optical pumping with intensity~$P_1(t,\mathbf{r})$.
$g _{1}$ is the constant of interaction of polaritons with reservoir excitons, $R_1$ is the stimulated scattering rate from the reservoir to the condensate, $\gamma _{\text{R}}$ is the exciton decay rate.
The parameter $G_1$ characterizes repulsion from the dark exciton reservoir, which also inevitably emerges within the pump spot.

The third term in Eq.~\eqref{EqPotential} is responsible for the potential, arising from the signal pulse of intensity~$P_2(t,\mathbf{r})$:
\begin{equation}
\label{EqPotentialin} U_{\text{in}}(t,\mathbf{r}) = \left(\frac{g_{2} + i R_2}{2(\gamma_{\text{R}} + R_2 |\Psi (t,\mathbf{r})|^2)} + G_2 \right) P_2 (t,\mathbf{r}).
\end{equation}
In Ref.~\cite{PhysRevLett124207402}, it is suggested to use cross-circularly polarized pump $P_2(t,\mathbf{r})$ in relation to $P_1(t,\mathbf{r})$.
This approach aims to reduce gain of the condensates in the dyad due to their overlapping with the signal.
The parameters in Eq.~\eqref{EqPotentialin} for $U_{\text{in}}(t,\mathbf{r})$ have the same meaning as those in Eq.~\eqref{EqPotentiald} for~$U_{\text{d}}(t,\mathbf{r})$.
We take the pumps in a Gaussian form as
\begin{subequations}
\begin{eqnarray}
&& P_{1} (t,\mathbf{r}) = P_{10} \exp \left( -t^2 / w_{\tau 1}^2 \right) \sum _{j,k} \exp \left[ -(x + j d)^2 / 2 w_1^2 + (y + k d)^2 / 2 w_1^2 \right], \\
&& P_{2} (t,\mathbf{r}) = \exp \left( -t^2 / w_{\tau 2}^2 \right) \sum _{j,k} P_{20} ^{(j,k)}\exp \left\{ -[x + (j+0.5) d]^2 / 2 w_2^2 + [y + (k+0.5) d]^2 / 2 w_2^2 \right\},
\end{eqnarray}
\end{subequations}
where $w_{\tau1, \tau2}$ and $w_{1,2}$ are durations and spatial widths of the pump pulses, $d$ is the distance between the centers of the pump spots.
The amplitude $P_{10}$ is taken equal for all pulses across the lattice unless otherwise indicated.
The amplitudes ${P_{20} ^{(j,k)} = P_{20}}$ for `1' elements of the signal lattice and ${P_{20} ^{(j,k)} = 0}$ for `0' elements.

The last term in Eq.~\eqref{EqPotential}, which differs our model from one in Ref.~\cite{PhysRevLett124207402}, characterizes the stationary barrier.
The barrier $V (\mathbf{r}) = F(\mathbf{r}, V_0, u, a) $, schematically shown in Fig.~\ref{FIG_ManyDyads}(c), is a function of a spatial coordinate, with the following parameters: $V_0$ is the height of the potential, $u$ is the  width of the walls, and $a$ is the width of the gap in the wall, responsible for entrance of the input signal.
Remarkably, the magnitude $V_0$ can be both real and imaginary, depending on the nature of the barrier.

%SSSSSSSSSSSSSSSSS
%SSSSSSSSSSSSSSSSS
%EEEEEEEEEEEEEEEEE
%CCCCCCCCCCCCCCCCC
{
\subsection*{Settings for simulation parameters}
}
For simulation, we take the following parameters suggested in~Ref.~\cite{PhysRevLett124207402}.
The effective polariton mass is ${m^* = 0.49 \, \text{meV ps}^2 \mu\text{m}^{-2}}$, 
the decay rates are ${\gamma = 1/6 \, \text{ps}^{-1}}$ and ${\gamma _{\text{R}} = 0.05 \, \text{ps}^{-1}}$,
the interaction constants are ${\alpha = 2.4 \, \mu \text{eV} \, \mu \text{m}^2}$, ${g_1 = 0.8 \alpha}$ and ${g_2 = 1.8 \alpha }$,
repulsion constants from dark reservoirs are $G_{1,2} = 4g_{1,2}/\gamma _{\text{R}}$,
the durations and spatial widths of the pump pulses are ${w_{\tau1} = 5}$~ps, ${w_{\tau2} = 8}$~ps and ${w_{1,2} = 2.2 \, \mu \text{m}}$, the distance between the pump spots is~${d = 12 \, \mu\text{m}}$.
We also take the scattering rates as ${R_1 = 7 \alpha}$ and ${R_2 = 0.7 R_1}$.  \\

%SSSSSSSSSSSSSSSSS
%SSSSSSSSSSSSSSSSS
%EEEEEEEEEEEEEEEEE
%CCCCCCCCCCCCCCCCC
{
\subsection*{Preprocessing of Speech Command audio data}

The Speech Commands dataset~\cite{warden2018speechcommandsdatasetlimitedvocabulary,speechcommandsv2} comprises WAV audio files that are preprocessed through a defined sequence to make them suitable for the neuromorphic network analysis:

\begin{itemize}
\item  \textit{Audio Length Adjustment}: While each audio file in the Speech Commands dataset is intended to be approximately one second long, some are shorter due to truncated recordings, often cutting off the very words they are meant to capture.
We remove these incomplete files from our dataset to maintain consistency and quality, only using full-length recordings for our analysis.

\item \textit{MFCC Extraction}: We compute MFCCs from each audio file, extracting 12 coefficients per frame across 24 frames.
This configuration balances detailed audio analysis with computational efficiency, ensuring we capture essential speech characteristics effectively for each one-second clip.
See Supplementary Section~5 for more details about~MFCC feature matrix.

\item \textit{Binarization}: The MFCC feature arrays are binarized.
Non-positive values are set to 0 and positive values are set to 1.

\item \textit{Dataset Splitting}: The processed data is split into training and testing sets using a 90:10 ratio, ensuring both effective training and thorough performance evaluation.
\end{itemize}

}

%BBBBBBBBBBBBBBBBB
%IIIIIIIIIIIIIIIII
%BBBBBBBBBBBBBBBBB
%LLLLLLLLLLLLLLLLL
\bibliographystyle{apsrev4-1}
\bibliography{polaritonBrainBibl}

\begin{thebibliography}{62}%
\makeatletter
\providecommand \@ifxundefined [1]{%
 \@ifx{#1\undefined}
}%
\providecommand \@ifnum [1]{%
 \ifnum #1\expandafter \@firstoftwo
 \else \expandafter \@secondoftwo
 \fi
}%
\providecommand \@ifx [1]{%
 \ifx #1\expandafter \@firstoftwo
 \else \expandafter \@secondoftwo
 \fi
}%
\providecommand \natexlab [1]{#1}%
\providecommand \enquote  [1]{``#1''}%
\providecommand \bibnamefont  [1]{#1}%
\providecommand \bibfnamefont [1]{#1}%
\providecommand \citenamefont [1]{#1}%
\providecommand \href@noop [0]{\@secondoftwo}%
\providecommand \href [0]{\begingroup \@sanitize@url \@href}%
\providecommand \@href[1]{\@@startlink{#1}\@@href}%
\providecommand \@@href[1]{\endgroup#1\@@endlink}%
\providecommand \@sanitize@url [0]{\catcode `\\12\catcode `\$12\catcode
  `\&12\catcode `\#12\catcode `\^12\catcode `\_12\catcode `\%12\relax}%
\providecommand \@@startlink[1]{}%
\providecommand \@@endlink[0]{}%
\providecommand \url  [0]{\begingroup\@sanitize@url \@url }%
\providecommand \@url [1]{\endgroup\@href {#1}{\urlprefix }}%
\providecommand \urlprefix  [0]{URL }%
\providecommand \Eprint [0]{\href }%
\providecommand \doibase [0]{http://dx.doi.org/}%
\providecommand \selectlanguage [0]{\@gobble}%
\providecommand \bibinfo  [0]{\@secondoftwo}%
\providecommand \bibfield  [0]{\@secondoftwo}%
\providecommand \translation [1]{[#1]}%
\providecommand \BibitemOpen [0]{}%
\providecommand \bibitemStop [0]{}%
\providecommand \bibitemNoStop [0]{.\EOS\space}%
\providecommand \EOS [0]{\spacefactor3000\relax}%
\providecommand \BibitemShut  [1]{\csname bibitem#1\endcsname}%
\let\auto@bib@innerbib\@empty
%</preamble>
\bibitem [{\citenamefont {Zhu}\ \emph {et~al.}(2020)\citenamefont {Zhu},
  \citenamefont {Zhang}, \citenamefont {Yang},\ and\ \citenamefont
  {Huang}}]{APRes7011312}%
  \BibitemOpen
  \bibfield  {author} {\bibinfo {author} {\bibfnamefont {J.}~\bibnamefont
  {Zhu}}, \bibinfo {author} {\bibfnamefont {T.}~\bibnamefont {Zhang}}, \bibinfo
  {author} {\bibfnamefont {Y.}~\bibnamefont {Yang}}, \ and\ \bibinfo {author}
  {\bibfnamefont {R.}~\bibnamefont {Huang}},\ }\href {\doibase
  10.1063/1.5118217} {\bibfield  {journal} {\bibinfo  {journal} {Applied
  Physics Reviews}\ }\textbf {\bibinfo {volume} {7}},\ \bibinfo {pages}
  {011312} (\bibinfo {year} {2020})}\BibitemShut {NoStop}%
\bibitem [{\citenamefont {Ostrau}\ \emph {et~al.}(2022)\citenamefont {Ostrau},
  \citenamefont {Klarhorst}, \citenamefont {Thies},\ and\ \citenamefont
  {Rückert}}]{FrontNeur16453X}%
  \BibitemOpen
  \bibfield  {author} {\bibinfo {author} {\bibfnamefont {C.}~\bibnamefont
  {Ostrau}}, \bibinfo {author} {\bibfnamefont {C.}~\bibnamefont {Klarhorst}},
  \bibinfo {author} {\bibfnamefont {M.}~\bibnamefont {Thies}}, \ and\ \bibinfo
  {author} {\bibfnamefont {U.}~\bibnamefont {Rückert}},\ }\href {\doibase
  10.3389/fnins.2022.873935} {\bibfield  {journal} {\bibinfo  {journal}
  {Frontiers in Neuroscience}\ }\textbf {\bibinfo {volume} {16}} (\bibinfo
  {year} {2022}),\ 10.3389/fnins.2022.873935}\BibitemShut {NoStop}%
\bibitem [{\citenamefont {Sangwan}\ and\ \citenamefont
  {Hersam}(2020)}]{NatMat15517}%
  \BibitemOpen
  \bibfield  {author} {\bibinfo {author} {\bibfnamefont {V.~K.}\ \bibnamefont
  {Sangwan}}\ and\ \bibinfo {author} {\bibfnamefont {M.~C.}\ \bibnamefont
  {Hersam}},\ }\href {\doibase 10.1038/s41565-020-0647-z} {\bibfield  {journal}
  {\bibinfo  {journal} {Nature Nanotechnology}\ }\textbf {\bibinfo {volume}
  {15}},\ \bibinfo {pages} {517} (\bibinfo {year} {2020})}\BibitemShut
  {NoStop}%
\bibitem [{\citenamefont {Ballarini}\ \emph {et~al.}(2020)\citenamefont
  {Ballarini}, \citenamefont {Gianfrate}, \citenamefont {Panico}, \citenamefont
  {Opala}, \citenamefont {Ghosh}, \citenamefont {Dominici}, \citenamefont
  {Ardizzone}, \citenamefont {De~Giorgi}, \citenamefont {Lerario},
  \citenamefont {Gigli}, \citenamefont {Liew}, \citenamefont {Matuszewski},\
  and\ \citenamefont {Sanvitto}}]{Nanolett203506}%
  \BibitemOpen
  \bibfield  {author} {\bibinfo {author} {\bibfnamefont {D.}~\bibnamefont
  {Ballarini}}, \bibinfo {author} {\bibfnamefont {A.}~\bibnamefont
  {Gianfrate}}, \bibinfo {author} {\bibfnamefont {R.}~\bibnamefont {Panico}},
  \bibinfo {author} {\bibfnamefont {A.}~\bibnamefont {Opala}}, \bibinfo
  {author} {\bibfnamefont {S.}~\bibnamefont {Ghosh}}, \bibinfo {author}
  {\bibfnamefont {L.}~\bibnamefont {Dominici}}, \bibinfo {author}
  {\bibfnamefont {V.}~\bibnamefont {Ardizzone}}, \bibinfo {author}
  {\bibfnamefont {M.}~\bibnamefont {De~Giorgi}}, \bibinfo {author}
  {\bibfnamefont {G.}~\bibnamefont {Lerario}}, \bibinfo {author} {\bibfnamefont
  {G.}~\bibnamefont {Gigli}}, \bibinfo {author} {\bibfnamefont {T.~C.~H.}\
  \bibnamefont {Liew}}, \bibinfo {author} {\bibfnamefont {M.}~\bibnamefont
  {Matuszewski}}, \ and\ \bibinfo {author} {\bibfnamefont {D.}~\bibnamefont
  {Sanvitto}},\ }\href {\doibase 10.1021/acs.nanolett.0c00435} {\bibfield
  {journal} {\bibinfo  {journal} {Nano Letters}\ }\textbf {\bibinfo {volume}
  {20}},\ \bibinfo {pages} {3506} (\bibinfo {year} {2020})}\BibitemShut
  {NoStop}%
\bibitem [{\citenamefont {Opala}\ \emph {et~al.}(2022)\citenamefont {Opala},
  \citenamefont {Panico}, \citenamefont {Ardizzone}, \citenamefont {Pietka},
  \citenamefont {Szczytko}, \citenamefont {Sanvitto}, \citenamefont
  {Matuszewski},\ and\ \citenamefont {Ballarini}}]{PhysRevApplied18024028}%
  \BibitemOpen
  \bibfield  {author} {\bibinfo {author} {\bibfnamefont {A.}~\bibnamefont
  {Opala}}, \bibinfo {author} {\bibfnamefont {R.}~\bibnamefont {Panico}},
  \bibinfo {author} {\bibfnamefont {V.}~\bibnamefont {Ardizzone}}, \bibinfo
  {author} {\bibfnamefont {B.}~\bibnamefont {Pietka}}, \bibinfo {author}
  {\bibfnamefont {J.}~\bibnamefont {Szczytko}}, \bibinfo {author}
  {\bibfnamefont {D.}~\bibnamefont {Sanvitto}}, \bibinfo {author}
  {\bibfnamefont {M.}~\bibnamefont {Matuszewski}}, \ and\ \bibinfo {author}
  {\bibfnamefont {D.}~\bibnamefont {Ballarini}},\ }\href {\doibase
  10.1103/PhysRevApplied.18.024028} {\bibfield  {journal} {\bibinfo  {journal}
  {Phys. Rev. Appl.}\ }\textbf {\bibinfo {volume} {18}},\ \bibinfo {pages}
  {024028} (\bibinfo {year} {2022})}\BibitemShut {NoStop}%
\bibitem [{\citenamefont {Kavokin}\ \emph {et~al.}(2017)\citenamefont
  {Kavokin}, \citenamefont {Baumberg}, \citenamefont {Malpuech},\ and\
  \citenamefont {Laussy}}]{kavokinBook2017}%
  \BibitemOpen
  \bibfield  {author} {\bibinfo {author} {\bibfnamefont {A.}~\bibnamefont
  {Kavokin}}, \bibinfo {author} {\bibfnamefont {J.}~\bibnamefont {Baumberg}},
  \bibinfo {author} {\bibfnamefont {G.}~\bibnamefont {Malpuech}}, \ and\
  \bibinfo {author} {\bibfnamefont {F.}~\bibnamefont {Laussy}},\ }\href@noop {}
  {\emph {\bibinfo {title} {Microcavities}}},\ \bibinfo {edition} {2nd}\ ed.,\
  Series on Semiconductor Science and Technology\ (\bibinfo  {publisher} {OUP
  Oxford},\ \bibinfo {year} {2017})\BibitemShut {NoStop}%
\bibitem [{\citenamefont {Carusotto}\ and\ \citenamefont
  {Ciuti}(2013)}]{RevModPhys85299}%
  \BibitemOpen
  \bibfield  {author} {\bibinfo {author} {\bibfnamefont {I.}~\bibnamefont
  {Carusotto}}\ and\ \bibinfo {author} {\bibfnamefont {C.}~\bibnamefont
  {Ciuti}},\ }\href {\doibase 10.1103/RevModPhys.85.299} {\bibfield  {journal}
  {\bibinfo  {journal} {Rev. Mod. Phys.}\ }\textbf {\bibinfo {volume} {85}},\
  \bibinfo {pages} {299} (\bibinfo {year} {2013})}\BibitemShut {NoStop}%
\bibitem [{\citenamefont {Liew}\ \emph {et~al.}(2008)\citenamefont {Liew},
  \citenamefont {Kavokin},\ and\ \citenamefont
  {Shelykh}}]{PhysRevLett101016402}%
  \BibitemOpen
  \bibfield  {author} {\bibinfo {author} {\bibfnamefont {T.~C.~H.}\
  \bibnamefont {Liew}}, \bibinfo {author} {\bibfnamefont {A.~V.}\ \bibnamefont
  {Kavokin}}, \ and\ \bibinfo {author} {\bibfnamefont {I.~A.}\ \bibnamefont
  {Shelykh}},\ }\href {\doibase 10.1103/PhysRevLett.101.016402} {\bibfield
  {journal} {\bibinfo  {journal} {Phys. Rev. Lett.}\ }\textbf {\bibinfo
  {volume} {101}},\ \bibinfo {pages} {016402} (\bibinfo {year}
  {2008})}\BibitemShut {NoStop}%
\bibitem [{\citenamefont {Opala}\ \emph {et~al.}(2019)\citenamefont {Opala},
  \citenamefont {Ghosh}, \citenamefont {Liew},\ and\ \citenamefont
  {Matuszewski}}]{PhysRevApplied11064029}%
  \BibitemOpen
  \bibfield  {author} {\bibinfo {author} {\bibfnamefont {A.}~\bibnamefont
  {Opala}}, \bibinfo {author} {\bibfnamefont {S.}~\bibnamefont {Ghosh}},
  \bibinfo {author} {\bibfnamefont {T.~C.}\ \bibnamefont {Liew}}, \ and\
  \bibinfo {author} {\bibfnamefont {M.}~\bibnamefont {Matuszewski}},\ }\href
  {\doibase 10.1103/PhysRevApplied.11.064029} {\bibfield  {journal} {\bibinfo
  {journal} {Phys. Rev. Appl.}\ }\textbf {\bibinfo {volume} {11}},\ \bibinfo
  {pages} {064029} (\bibinfo {year} {2019})}\BibitemShut {NoStop}%
\bibitem [{\citenamefont {Xu}\ \emph {et~al.}(2020)\citenamefont {Xu},
  \citenamefont {Ghosh}, \citenamefont {Matuszewski},\ and\ \citenamefont
  {Liew}}]{PhysRevApplied13064074}%
  \BibitemOpen
  \bibfield  {author} {\bibinfo {author} {\bibfnamefont {H.}~\bibnamefont
  {Xu}}, \bibinfo {author} {\bibfnamefont {S.}~\bibnamefont {Ghosh}}, \bibinfo
  {author} {\bibfnamefont {M.}~\bibnamefont {Matuszewski}}, \ and\ \bibinfo
  {author} {\bibfnamefont {T.~C.}\ \bibnamefont {Liew}},\ }\href {\doibase
  10.1103/PhysRevApplied.13.064074} {\bibfield  {journal} {\bibinfo  {journal}
  {Phys. Rev. Appl.}\ }\textbf {\bibinfo {volume} {13}},\ \bibinfo {pages}
  {064074} (\bibinfo {year} {2020})}\BibitemShut {NoStop}%
\bibitem [{\citenamefont {Ghosh}\ \emph {et~al.}(2019)\citenamefont {Ghosh},
  \citenamefont {Paterek},\ and\ \citenamefont {Liew}}]{PhysRevLett123260404}%
  \BibitemOpen
  \bibfield  {author} {\bibinfo {author} {\bibfnamefont {S.}~\bibnamefont
  {Ghosh}}, \bibinfo {author} {\bibfnamefont {T.}~\bibnamefont {Paterek}}, \
  and\ \bibinfo {author} {\bibfnamefont {T.~C.~H.}\ \bibnamefont {Liew}},\
  }\href {\doibase 10.1103/PhysRevLett.123.260404} {\bibfield  {journal}
  {\bibinfo  {journal} {Phys. Rev. Lett.}\ }\textbf {\bibinfo {volume} {123}},\
  \bibinfo {pages} {260404} (\bibinfo {year} {2019})}\BibitemShut {NoStop}%
\bibitem [{\citenamefont {{LeCun, Yann and Bottou, L{\'e}on and Bengio, Yoshua
  and Haffner, Patrick}}(1998)}]{lecun1998gradient}%
  \BibitemOpen
  \bibfield  {author} {\bibinfo {author} {\bibnamefont {{LeCun, Yann and
  Bottou, L{\'e}on and Bengio, Yoshua and Haffner, Patrick}}},\ }\href@noop {}
  {\bibfield  {journal} {\bibinfo  {journal} {Proceedings of the IEEE}\
  }\textbf {\bibinfo {volume} {86}},\ \bibinfo {pages} {2278} (\bibinfo {year}
  {1998})}\BibitemShut {NoStop}%
\bibitem [{\citenamefont {{LeCun, Yann and Cortes, Corinna and Burges, C
  J}}(2010)}]{lecun2010mnist}%
  \BibitemOpen
  \bibfield  {author} {\bibinfo {author} {\bibnamefont {{LeCun, Yann and
  Cortes, Corinna and Burges, C J}}},\ }\href@noop {} {\bibfield  {journal}
  {\bibinfo  {journal} {ATT Labs [Online]. Available:
  \url{http://yann.lecun.com/exdb/mnist}}\ }\textbf {\bibinfo {volume} {2}}
  (\bibinfo {year} {2010})}\BibitemShut {NoStop}%
\bibitem [{\citenamefont {Mehonic}\ and\ \citenamefont
  {Kenyon}(2022)}]{Nature604255}%
  \BibitemOpen
  \bibfield  {author} {\bibinfo {author} {\bibfnamefont {A.}~\bibnamefont
  {Mehonic}}\ and\ \bibinfo {author} {\bibfnamefont {A.~J.}\ \bibnamefont
  {Kenyon}},\ }\href {\doibase 10.1038/s41586-021-04362-w} {\bibfield
  {journal} {\bibinfo  {journal} {Nature}\ }\textbf {\bibinfo {volume} {604}},\
  \bibinfo {pages} {255} (\bibinfo {year} {2022})}\BibitemShut {NoStop}%
\bibitem [{\citenamefont {Mirek}\ \emph {et~al.}(2021)\citenamefont {Mirek},
  \citenamefont {Opala}, \citenamefont {Comaron}, \citenamefont {Furman},
  \citenamefont {Kr{\'o}l}, \citenamefont {Tyszka}, \citenamefont
  {Seredy{\'n}ski}, \citenamefont {Ballarini}, \citenamefont {Sanvitto},
  \citenamefont {Liew}, \citenamefont {Pacuski}, \citenamefont
  {Suffczy{\'n}ski}, \citenamefont {Szczytko}, \citenamefont {Matuszewski},\
  and\ \citenamefont {Pietka}}]{NanoLett213715}%
  \BibitemOpen
  \bibfield  {author} {\bibinfo {author} {\bibfnamefont {R.}~\bibnamefont
  {Mirek}}, \bibinfo {author} {\bibfnamefont {A.}~\bibnamefont {Opala}},
  \bibinfo {author} {\bibfnamefont {P.}~\bibnamefont {Comaron}}, \bibinfo
  {author} {\bibfnamefont {M.}~\bibnamefont {Furman}}, \bibinfo {author}
  {\bibfnamefont {M.}~\bibnamefont {Kr{\'o}l}}, \bibinfo {author}
  {\bibfnamefont {K.}~\bibnamefont {Tyszka}}, \bibinfo {author} {\bibfnamefont
  {B.}~\bibnamefont {Seredy{\'n}ski}}, \bibinfo {author} {\bibfnamefont
  {D.}~\bibnamefont {Ballarini}}, \bibinfo {author} {\bibfnamefont
  {D.}~\bibnamefont {Sanvitto}}, \bibinfo {author} {\bibfnamefont {T.~C.~H.}\
  \bibnamefont {Liew}}, \bibinfo {author} {\bibfnamefont {W.}~\bibnamefont
  {Pacuski}}, \bibinfo {author} {\bibfnamefont {J.}~\bibnamefont
  {Suffczy{\'n}ski}}, \bibinfo {author} {\bibfnamefont {J.}~\bibnamefont
  {Szczytko}}, \bibinfo {author} {\bibfnamefont {M.}~\bibnamefont
  {Matuszewski}}, \ and\ \bibinfo {author} {\bibfnamefont {B.}~\bibnamefont
  {Pietka}},\ }\href {\doibase 10.1021/acs.nanolett.0c04696} {\bibfield
  {journal} {\bibinfo  {journal} {Nano Letters}\ }\textbf {\bibinfo {volume}
  {21}},\ \bibinfo {pages} {3715} (\bibinfo {year} {2021})}\BibitemShut
  {NoStop}%
\bibitem [{\citenamefont {Matuszewski}\ \emph {et~al.}(2021)\citenamefont
  {Matuszewski}, \citenamefont {Opala}, \citenamefont {Mirek}, \citenamefont
  {Furman}, \citenamefont {Kr\'ol}, \citenamefont {Tyszka}, \citenamefont
  {Liew}, \citenamefont {Ballarini}, \citenamefont {Sanvitto}, \citenamefont
  {Szczytko},\ and\ \citenamefont {Pietka}}]{PhysRevApplied16024045}%
  \BibitemOpen
  \bibfield  {author} {\bibinfo {author} {\bibfnamefont {M.}~\bibnamefont
  {Matuszewski}}, \bibinfo {author} {\bibfnamefont {A.}~\bibnamefont {Opala}},
  \bibinfo {author} {\bibfnamefont {R.}~\bibnamefont {Mirek}}, \bibinfo
  {author} {\bibfnamefont {M.}~\bibnamefont {Furman}}, \bibinfo {author}
  {\bibfnamefont {M.}~\bibnamefont {Kr\'ol}}, \bibinfo {author} {\bibfnamefont
  {K.}~\bibnamefont {Tyszka}}, \bibinfo {author} {\bibfnamefont
  {T.}~\bibnamefont {Liew}}, \bibinfo {author} {\bibfnamefont {D.}~\bibnamefont
  {Ballarini}}, \bibinfo {author} {\bibfnamefont {D.}~\bibnamefont {Sanvitto}},
  \bibinfo {author} {\bibfnamefont {J.}~\bibnamefont {Szczytko}}, \ and\
  \bibinfo {author} {\bibfnamefont {B.}~\bibnamefont {Pietka}},\ }\href
  {\doibase 10.1103/PhysRevApplied.16.024045} {\bibfield  {journal} {\bibinfo
  {journal} {Phys. Rev. Appl.}\ }\textbf {\bibinfo {volume} {16}},\ \bibinfo
  {pages} {024045} (\bibinfo {year} {2021})}\BibitemShut {NoStop}%
\bibitem [{\citenamefont {Bloch}\ \emph {et~al.}(2008)\citenamefont {Bloch},
  \citenamefont {Dalibard},\ and\ \citenamefont {Zwerger}}]{RevModPhys80885}%
  \BibitemOpen
  \bibfield  {author} {\bibinfo {author} {\bibfnamefont {I.}~\bibnamefont
  {Bloch}}, \bibinfo {author} {\bibfnamefont {J.}~\bibnamefont {Dalibard}}, \
  and\ \bibinfo {author} {\bibfnamefont {W.}~\bibnamefont {Zwerger}},\ }\href
  {\doibase 10.1103/RevModPhys.80.885} {\bibfield  {journal} {\bibinfo
  {journal} {Rev. Mod. Phys.}\ }\textbf {\bibinfo {volume} {80}},\ \bibinfo
  {pages} {885} (\bibinfo {year} {2008})}\BibitemShut {NoStop}%
\bibitem [{\citenamefont {Greiner}\ \emph {et~al.}(2002)\citenamefont
  {Greiner}, \citenamefont {Mandel}, \citenamefont {Esslinger}, \citenamefont
  {H{\"a}nsch},\ and\ \citenamefont {Bloch}}]{Nature41539}%
  \BibitemOpen
  \bibfield  {author} {\bibinfo {author} {\bibfnamefont {M.}~\bibnamefont
  {Greiner}}, \bibinfo {author} {\bibfnamefont {O.}~\bibnamefont {Mandel}},
  \bibinfo {author} {\bibfnamefont {T.}~\bibnamefont {Esslinger}}, \bibinfo
  {author} {\bibfnamefont {T.~W.}\ \bibnamefont {H{\"a}nsch}}, \ and\ \bibinfo
  {author} {\bibfnamefont {I.}~\bibnamefont {Bloch}},\ }\href {\doibase
  10.1038/415039a} {\bibfield  {journal} {\bibinfo  {journal} {Nature}\
  }\textbf {\bibinfo {volume} {415}},\ \bibinfo {pages} {39} (\bibinfo {year}
  {2002})}\BibitemShut {NoStop}%
\bibitem [{\citenamefont {Chen}\ \emph {et~al.}(2012)\citenamefont {Chen},
  \citenamefont {Lin}, \citenamefont {Lai}, \citenamefont {Sedov},
  \citenamefont {Alodjants}, \citenamefont {Arakelian},\ and\ \citenamefont
  {Lee}}]{PhysRevA86023829}%
  \BibitemOpen
  \bibfield  {author} {\bibinfo {author} {\bibfnamefont {I.-H.}\ \bibnamefont
  {Chen}}, \bibinfo {author} {\bibfnamefont {Y.~Y.}\ \bibnamefont {Lin}},
  \bibinfo {author} {\bibfnamefont {Y.-C.}\ \bibnamefont {Lai}}, \bibinfo
  {author} {\bibfnamefont {E.~S.}\ \bibnamefont {Sedov}}, \bibinfo {author}
  {\bibfnamefont {A.~P.}\ \bibnamefont {Alodjants}}, \bibinfo {author}
  {\bibfnamefont {S.~M.}\ \bibnamefont {Arakelian}}, \ and\ \bibinfo {author}
  {\bibfnamefont {R.-K.}\ \bibnamefont {Lee}},\ }\href {\doibase
  10.1103/PhysRevA.86.023829} {\bibfield  {journal} {\bibinfo  {journal} {Phys.
  Rev. A}\ }\textbf {\bibinfo {volume} {86}},\ \bibinfo {pages} {023829}
  (\bibinfo {year} {2012})}\BibitemShut {NoStop}%
\bibitem [{\citenamefont {Sedov}\ \emph {et~al.}(2014)\citenamefont {Sedov},
  \citenamefont {Alodjants}, \citenamefont {Arakelian}, \citenamefont {Chuang},
  \citenamefont {Lin}, \citenamefont {Yang},\ and\ \citenamefont
  {Lee}}]{PhysRevA89033828}%
  \BibitemOpen
  \bibfield  {author} {\bibinfo {author} {\bibfnamefont {E.~S.}\ \bibnamefont
  {Sedov}}, \bibinfo {author} {\bibfnamefont {A.~P.}\ \bibnamefont
  {Alodjants}}, \bibinfo {author} {\bibfnamefont {S.~M.}\ \bibnamefont
  {Arakelian}}, \bibinfo {author} {\bibfnamefont {Y.-L.}\ \bibnamefont
  {Chuang}}, \bibinfo {author} {\bibfnamefont {Y.}~\bibnamefont {Lin}},
  \bibinfo {author} {\bibfnamefont {W.-X.}\ \bibnamefont {Yang}}, \ and\
  \bibinfo {author} {\bibfnamefont {R.-K.}\ \bibnamefont {Lee}},\ }\href
  {\doibase 10.1103/PhysRevA.89.033828} {\bibfield  {journal} {\bibinfo
  {journal} {Phys. Rev. A}\ }\textbf {\bibinfo {volume} {89}},\ \bibinfo
  {pages} {033828} (\bibinfo {year} {2014})}\BibitemShut {NoStop}%
\bibitem [{\citenamefont {Kwon}\ \emph {et~al.}(2022)\citenamefont {Kwon},
  \citenamefont {Kim}, \citenamefont {Lanuza},\ and\ \citenamefont
  {Schneble}}]{NatPhys18657}%
  \BibitemOpen
  \bibfield  {author} {\bibinfo {author} {\bibfnamefont {J.}~\bibnamefont
  {Kwon}}, \bibinfo {author} {\bibfnamefont {Y.}~\bibnamefont {Kim}}, \bibinfo
  {author} {\bibfnamefont {A.}~\bibnamefont {Lanuza}}, \ and\ \bibinfo {author}
  {\bibfnamefont {D.}~\bibnamefont {Schneble}},\ }\href {\doibase
  10.1038/s41567-022-01565-4} {\bibfield  {journal} {\bibinfo  {journal}
  {Nature Physics}\ }\textbf {\bibinfo {volume} {18}},\ \bibinfo {pages} {657}
  (\bibinfo {year} {2022})}\BibitemShut {NoStop}%
\bibitem [{\citenamefont {Amo}\ and\ \citenamefont
  {Bloch}(2016)}]{CompRenPhys17934}%
  \BibitemOpen
  \bibfield  {author} {\bibinfo {author} {\bibfnamefont {A.}~\bibnamefont
  {Amo}}\ and\ \bibinfo {author} {\bibfnamefont {J.}~\bibnamefont {Bloch}},\
  }\href {\doibase https://doi.org/10.1016/j.crhy.2016.08.007} {\bibfield
  {journal} {\bibinfo  {journal} {Comptes Rendus Physique}\ }\textbf {\bibinfo
  {volume} {17}},\ \bibinfo {pages} {934} (\bibinfo {year} {2016})},\ \bibinfo
  {note} {polariton physics / Physique des polaritons}\BibitemShut {NoStop}%
\bibitem [{\citenamefont {Kavokin}\ \emph {et~al.}(2022)\citenamefont
  {Kavokin}, \citenamefont {Liew}, \citenamefont {Schneider}, \citenamefont
  {Lagoudakis}, \citenamefont {Klembt},\ and\ \citenamefont
  {Hoefling}}]{NatRevPhys4435}%
  \BibitemOpen
  \bibfield  {author} {\bibinfo {author} {\bibfnamefont {A.}~\bibnamefont
  {Kavokin}}, \bibinfo {author} {\bibfnamefont {T.~C.~H.}\ \bibnamefont
  {Liew}}, \bibinfo {author} {\bibfnamefont {C.}~\bibnamefont {Schneider}},
  \bibinfo {author} {\bibfnamefont {P.~G.}\ \bibnamefont {Lagoudakis}},
  \bibinfo {author} {\bibfnamefont {S.}~\bibnamefont {Klembt}}, \ and\ \bibinfo
  {author} {\bibfnamefont {S.}~\bibnamefont {Hoefling}},\ }\href {\doibase
  10.1038/s42254-022-00447-1} {\bibfield  {journal} {\bibinfo  {journal}
  {Nature Reviews Physics}\ }\textbf {\bibinfo {volume} {4}},\ \bibinfo {pages}
  {435} (\bibinfo {year} {2022})}\BibitemShut {NoStop}%
\bibitem [{\citenamefont {Lagoudakis}\ and\ \citenamefont
  {Berloff}(2017)}]{NJPhys19125008}%
  \BibitemOpen
  \bibfield  {author} {\bibinfo {author} {\bibfnamefont {P.~G.}\ \bibnamefont
  {Lagoudakis}}\ and\ \bibinfo {author} {\bibfnamefont {N.~G.}\ \bibnamefont
  {Berloff}},\ }\href {\doibase 10.1088/1367-2630/aa924b} {\bibfield  {journal}
  {\bibinfo  {journal} {New Journal of Physics}\ }\textbf {\bibinfo {volume}
  {19}},\ \bibinfo {pages} {125008} (\bibinfo {year} {2017})}\BibitemShut
  {NoStop}%
\bibitem [{\citenamefont {Berloff}\ \emph {et~al.}(2017)\citenamefont
  {Berloff}, \citenamefont {Silva}, \citenamefont {Kalinin}, \citenamefont
  {Askitopoulos}, \citenamefont {T{\"o}pfer}, \citenamefont {Cilibrizzi},
  \citenamefont {Langbein},\ and\ \citenamefont {Lagoudakis}}]{NatMater161120}%
  \BibitemOpen
  \bibfield  {author} {\bibinfo {author} {\bibfnamefont {N.~G.}\ \bibnamefont
  {Berloff}}, \bibinfo {author} {\bibfnamefont {M.}~\bibnamefont {Silva}},
  \bibinfo {author} {\bibfnamefont {K.}~\bibnamefont {Kalinin}}, \bibinfo
  {author} {\bibfnamefont {A.}~\bibnamefont {Askitopoulos}}, \bibinfo {author}
  {\bibfnamefont {J.~D.}\ \bibnamefont {T{\"o}pfer}}, \bibinfo {author}
  {\bibfnamefont {P.}~\bibnamefont {Cilibrizzi}}, \bibinfo {author}
  {\bibfnamefont {W.}~\bibnamefont {Langbein}}, \ and\ \bibinfo {author}
  {\bibfnamefont {P.~G.}\ \bibnamefont {Lagoudakis}},\ }\href {\doibase
  10.1038/nmat4971} {\bibfield  {journal} {\bibinfo  {journal} {Nature
  Materials}\ }\textbf {\bibinfo {volume} {16}},\ \bibinfo {pages} {1120}
  (\bibinfo {year} {2017})}\BibitemShut {NoStop}%
\bibitem [{\citenamefont {Whittaker}\ \emph {et~al.}(2018)\citenamefont
  {Whittaker}, \citenamefont {Cancellieri}, \citenamefont {Walker},
  \citenamefont {Gulevich}, \citenamefont {Schomerus}, \citenamefont
  {Vaitiekus}, \citenamefont {Royall}, \citenamefont {Whittaker}, \citenamefont
  {Clarke}, \citenamefont {Iorsh}, \citenamefont {Shelykh}, \citenamefont
  {Skolnick},\ and\ \citenamefont {Krizhanovskii}}]{PhysRevLett120097401}%
  \BibitemOpen
  \bibfield  {author} {\bibinfo {author} {\bibfnamefont {C.~E.}\ \bibnamefont
  {Whittaker}}, \bibinfo {author} {\bibfnamefont {E.}~\bibnamefont
  {Cancellieri}}, \bibinfo {author} {\bibfnamefont {P.~M.}\ \bibnamefont
  {Walker}}, \bibinfo {author} {\bibfnamefont {D.~R.}\ \bibnamefont
  {Gulevich}}, \bibinfo {author} {\bibfnamefont {H.}~\bibnamefont {Schomerus}},
  \bibinfo {author} {\bibfnamefont {D.}~\bibnamefont {Vaitiekus}}, \bibinfo
  {author} {\bibfnamefont {B.}~\bibnamefont {Royall}}, \bibinfo {author}
  {\bibfnamefont {D.~M.}\ \bibnamefont {Whittaker}}, \bibinfo {author}
  {\bibfnamefont {E.}~\bibnamefont {Clarke}}, \bibinfo {author} {\bibfnamefont
  {I.~V.}\ \bibnamefont {Iorsh}}, \bibinfo {author} {\bibfnamefont {I.~A.}\
  \bibnamefont {Shelykh}}, \bibinfo {author} {\bibfnamefont {M.~S.}\
  \bibnamefont {Skolnick}}, \ and\ \bibinfo {author} {\bibfnamefont {D.~N.}\
  \bibnamefont {Krizhanovskii}},\ }\href {\doibase
  10.1103/PhysRevLett.120.097401} {\bibfield  {journal} {\bibinfo  {journal}
  {Phys. Rev. Lett.}\ }\textbf {\bibinfo {volume} {120}},\ \bibinfo {pages}
  {097401} (\bibinfo {year} {2018})}\BibitemShut {NoStop}%
\bibitem [{\citenamefont {Kaitouni}\ \emph {et~al.}(2006)\citenamefont
  {Kaitouni}, \citenamefont {El~Da\"{\i}f}, \citenamefont {Baas}, \citenamefont
  {Richard}, \citenamefont {Paraiso}, \citenamefont {Lugan}, \citenamefont
  {Guillet}, \citenamefont {Morier-Genoud}, \citenamefont {Gani\`ere},
  \citenamefont {Staehli}, \citenamefont {Savona},\ and\ \citenamefont
  {Deveaud}}]{PhysRevB74155311}%
  \BibitemOpen
  \bibfield  {author} {\bibinfo {author} {\bibfnamefont {R.~I.}\ \bibnamefont
  {Kaitouni}}, \bibinfo {author} {\bibfnamefont {O.}~\bibnamefont
  {El~Da\"{\i}f}}, \bibinfo {author} {\bibfnamefont {A.}~\bibnamefont {Baas}},
  \bibinfo {author} {\bibfnamefont {M.}~\bibnamefont {Richard}}, \bibinfo
  {author} {\bibfnamefont {T.}~\bibnamefont {Paraiso}}, \bibinfo {author}
  {\bibfnamefont {P.}~\bibnamefont {Lugan}}, \bibinfo {author} {\bibfnamefont
  {T.}~\bibnamefont {Guillet}}, \bibinfo {author} {\bibfnamefont
  {F.}~\bibnamefont {Morier-Genoud}}, \bibinfo {author} {\bibfnamefont {J.~D.}\
  \bibnamefont {Gani\`ere}}, \bibinfo {author} {\bibfnamefont {J.~L.}\
  \bibnamefont {Staehli}}, \bibinfo {author} {\bibfnamefont {V.}~\bibnamefont
  {Savona}}, \ and\ \bibinfo {author} {\bibfnamefont {B.}~\bibnamefont
  {Deveaud}},\ }\href {\doibase 10.1103/PhysRevB.74.155311} {\bibfield
  {journal} {\bibinfo  {journal} {Phys. Rev. B}\ }\textbf {\bibinfo {volume}
  {74}},\ \bibinfo {pages} {155311} (\bibinfo {year} {2006})}\BibitemShut
  {NoStop}%
\bibitem [{\citenamefont {Kim}\ \emph {et~al.}(2011)\citenamefont {Kim},
  \citenamefont {Kusudo}, \citenamefont {Wu}, \citenamefont {Masumoto},
  \citenamefont {L{\"o}ffler}, \citenamefont {H{\"o}fling}, \citenamefont
  {Kumada}, \citenamefont {Worschech}, \citenamefont {Forchel},\ and\
  \citenamefont {Yamamoto}}]{NatPhys7681}%
  \BibitemOpen
  \bibfield  {author} {\bibinfo {author} {\bibfnamefont {N.~Y.}\ \bibnamefont
  {Kim}}, \bibinfo {author} {\bibfnamefont {K.}~\bibnamefont {Kusudo}},
  \bibinfo {author} {\bibfnamefont {C.}~\bibnamefont {Wu}}, \bibinfo {author}
  {\bibfnamefont {N.}~\bibnamefont {Masumoto}}, \bibinfo {author}
  {\bibfnamefont {A.}~\bibnamefont {L{\"o}ffler}}, \bibinfo {author}
  {\bibfnamefont {S.}~\bibnamefont {H{\"o}fling}}, \bibinfo {author}
  {\bibfnamefont {N.}~\bibnamefont {Kumada}}, \bibinfo {author} {\bibfnamefont
  {L.}~\bibnamefont {Worschech}}, \bibinfo {author} {\bibfnamefont
  {A.}~\bibnamefont {Forchel}}, \ and\ \bibinfo {author} {\bibfnamefont
  {Y.}~\bibnamefont {Yamamoto}},\ }\href {\doibase 10.1038/nphys2012}
  {\bibfield  {journal} {\bibinfo  {journal} {Nature Physics}\ }\textbf
  {\bibinfo {volume} {7}},\ \bibinfo {pages} {681} (\bibinfo {year}
  {2011})}\BibitemShut {NoStop}%
\bibitem [{\citenamefont {T\"{o}pfer}\ \emph {et~al.}(2021)\citenamefont
  {T\"{o}pfer}, \citenamefont {Chatzopoulos}, \citenamefont {Sigurdsson},
  \citenamefont {Cookson}, \citenamefont {Rubo},\ and\ \citenamefont
  {Lagoudakis}}]{Optica8106}%
  \BibitemOpen
  \bibfield  {author} {\bibinfo {author} {\bibfnamefont {J.~D.}\ \bibnamefont
  {T\"{o}pfer}}, \bibinfo {author} {\bibfnamefont {I.}~\bibnamefont
  {Chatzopoulos}}, \bibinfo {author} {\bibfnamefont {H.}~\bibnamefont
  {Sigurdsson}}, \bibinfo {author} {\bibfnamefont {T.}~\bibnamefont {Cookson}},
  \bibinfo {author} {\bibfnamefont {Y.~G.}\ \bibnamefont {Rubo}}, \ and\
  \bibinfo {author} {\bibfnamefont {P.~G.}\ \bibnamefont {Lagoudakis}},\ }\href
  {\doibase 10.1364/OPTICA.409976} {\bibfield  {journal} {\bibinfo  {journal}
  {Optica}\ }\textbf {\bibinfo {volume} {8}},\ \bibinfo {pages} {106} (\bibinfo
  {year} {2021})}\BibitemShut {NoStop}%
\bibitem [{\citenamefont {Alyatkin}\ \emph {et~al.}(2020)\citenamefont
  {Alyatkin}, \citenamefont {T\"opfer}, \citenamefont {Askitopoulos},
  \citenamefont {Sigurdsson},\ and\ \citenamefont
  {Lagoudakis}}]{PhysRevLett124207402}%
  \BibitemOpen
  \bibfield  {author} {\bibinfo {author} {\bibfnamefont {S.}~\bibnamefont
  {Alyatkin}}, \bibinfo {author} {\bibfnamefont {J.~D.}\ \bibnamefont
  {T\"opfer}}, \bibinfo {author} {\bibfnamefont {A.}~\bibnamefont
  {Askitopoulos}}, \bibinfo {author} {\bibfnamefont {H.}~\bibnamefont
  {Sigurdsson}}, \ and\ \bibinfo {author} {\bibfnamefont {P.~G.}\ \bibnamefont
  {Lagoudakis}},\ }\href {\doibase 10.1103/PhysRevLett.124.207402} {\bibfield
  {journal} {\bibinfo  {journal} {Phys. Rev. Lett.}\ }\textbf {\bibinfo
  {volume} {124}},\ \bibinfo {pages} {207402} (\bibinfo {year}
  {2020})}\BibitemShut {NoStop}%
\bibitem [{\citenamefont {Askitopoulos}\ \emph {et~al.}(2015)\citenamefont
  {Askitopoulos}, \citenamefont {Liew}, \citenamefont {Ohadi}, \citenamefont
  {Hatzopoulos}, \citenamefont {Savvidis},\ and\ \citenamefont
  {Lagoudakis}}]{PhysRevB92035305}%
  \BibitemOpen
  \bibfield  {author} {\bibinfo {author} {\bibfnamefont {A.}~\bibnamefont
  {Askitopoulos}}, \bibinfo {author} {\bibfnamefont {T.~C.~H.}\ \bibnamefont
  {Liew}}, \bibinfo {author} {\bibfnamefont {H.}~\bibnamefont {Ohadi}},
  \bibinfo {author} {\bibfnamefont {Z.}~\bibnamefont {Hatzopoulos}}, \bibinfo
  {author} {\bibfnamefont {P.~G.}\ \bibnamefont {Savvidis}}, \ and\ \bibinfo
  {author} {\bibfnamefont {P.~G.}\ \bibnamefont {Lagoudakis}},\ }\href
  {\doibase 10.1103/PhysRevB.92.035305} {\bibfield  {journal} {\bibinfo
  {journal} {Phys. Rev. B}\ }\textbf {\bibinfo {volume} {92}},\ \bibinfo
  {pages} {035305} (\bibinfo {year} {2015})}\BibitemShut {NoStop}%
\bibitem [{\citenamefont {Askitopoulos}\ \emph {et~al.}(2018)\citenamefont
  {Askitopoulos}, \citenamefont {Nalitov}, \citenamefont {Sedov}, \citenamefont
  {Pickup}, \citenamefont {Cherotchenko}, \citenamefont {Hatzopoulos},
  \citenamefont {Savvidis}, \citenamefont {Kavokin},\ and\ \citenamefont
  {Lagoudakis}}]{PhysRevB97235303}%
  \BibitemOpen
  \bibfield  {author} {\bibinfo {author} {\bibfnamefont {A.}~\bibnamefont
  {Askitopoulos}}, \bibinfo {author} {\bibfnamefont {A.~V.}\ \bibnamefont
  {Nalitov}}, \bibinfo {author} {\bibfnamefont {E.~S.}\ \bibnamefont {Sedov}},
  \bibinfo {author} {\bibfnamefont {L.}~\bibnamefont {Pickup}}, \bibinfo
  {author} {\bibfnamefont {E.~D.}\ \bibnamefont {Cherotchenko}}, \bibinfo
  {author} {\bibfnamefont {Z.}~\bibnamefont {Hatzopoulos}}, \bibinfo {author}
  {\bibfnamefont {P.~G.}\ \bibnamefont {Savvidis}}, \bibinfo {author}
  {\bibfnamefont {A.~V.}\ \bibnamefont {Kavokin}}, \ and\ \bibinfo {author}
  {\bibfnamefont {P.~G.}\ \bibnamefont {Lagoudakis}},\ }\href {\doibase
  10.1103/PhysRevB.97.235303} {\bibfield  {journal} {\bibinfo  {journal} {Phys.
  Rev. B}\ }\textbf {\bibinfo {volume} {97}},\ \bibinfo {pages} {235303}
  (\bibinfo {year} {2018})}\BibitemShut {NoStop}%
\bibitem [{\citenamefont {Vretenar}\ \emph {et~al.}(2021)\citenamefont
  {Vretenar}, \citenamefont {Kassenberg}, \citenamefont {Bissesar},
  \citenamefont {Toebes},\ and\ \citenamefont
  {Klaers}}]{PhysRevResearch3023167}%
  \BibitemOpen
  \bibfield  {author} {\bibinfo {author} {\bibfnamefont {M.}~\bibnamefont
  {Vretenar}}, \bibinfo {author} {\bibfnamefont {B.}~\bibnamefont
  {Kassenberg}}, \bibinfo {author} {\bibfnamefont {S.}~\bibnamefont
  {Bissesar}}, \bibinfo {author} {\bibfnamefont {C.}~\bibnamefont {Toebes}}, \
  and\ \bibinfo {author} {\bibfnamefont {J.}~\bibnamefont {Klaers}},\ }\href
  {\doibase 10.1103/PhysRevResearch.3.023167} {\bibfield  {journal} {\bibinfo
  {journal} {Phys. Rev. Res.}\ }\textbf {\bibinfo {volume} {3}},\ \bibinfo
  {pages} {023167} (\bibinfo {year} {2021})}\BibitemShut {NoStop}%
\bibitem [{\citenamefont {Lukoshkin}\ \emph {et~al.}(2018)\citenamefont
  {Lukoshkin}, \citenamefont {Kalevich}, \citenamefont {Afanasiev},
  \citenamefont {Kavokin}, \citenamefont {Hatzopoulos}, \citenamefont
  {Savvidis}, \citenamefont {Sedov},\ and\ \citenamefont
  {Kavokin}}]{PhysRevB97195149}%
  \BibitemOpen
  \bibfield  {author} {\bibinfo {author} {\bibfnamefont {V.~A.}\ \bibnamefont
  {Lukoshkin}}, \bibinfo {author} {\bibfnamefont {V.~K.}\ \bibnamefont
  {Kalevich}}, \bibinfo {author} {\bibfnamefont {M.~M.}\ \bibnamefont
  {Afanasiev}}, \bibinfo {author} {\bibfnamefont {K.~V.}\ \bibnamefont
  {Kavokin}}, \bibinfo {author} {\bibfnamefont {Z.}~\bibnamefont
  {Hatzopoulos}}, \bibinfo {author} {\bibfnamefont {P.~G.}\ \bibnamefont
  {Savvidis}}, \bibinfo {author} {\bibfnamefont {E.~S.}\ \bibnamefont {Sedov}},
  \ and\ \bibinfo {author} {\bibfnamefont {A.~V.}\ \bibnamefont {Kavokin}},\
  }\href {\doibase 10.1103/PhysRevB.97.195149} {\bibfield  {journal} {\bibinfo
  {journal} {Phys. Rev. B}\ }\textbf {\bibinfo {volume} {97}},\ \bibinfo
  {pages} {195149} (\bibinfo {year} {2018})}\BibitemShut {NoStop}%
\bibitem [{\citenamefont {Lukoshkin}\ \emph {et~al.}(2023)\citenamefont
  {Lukoshkin}, \citenamefont {Sedov}, \citenamefont {Kalevich}, \citenamefont
  {Hatzopoulos}, \citenamefont {Savvidis},\ and\ \citenamefont
  {Kavokin}}]{SciRep134607}%
  \BibitemOpen
  \bibfield  {author} {\bibinfo {author} {\bibfnamefont {V.}~\bibnamefont
  {Lukoshkin}}, \bibinfo {author} {\bibfnamefont {E.}~\bibnamefont {Sedov}},
  \bibinfo {author} {\bibfnamefont {V.}~\bibnamefont {Kalevich}}, \bibinfo
  {author} {\bibfnamefont {Z.}~\bibnamefont {Hatzopoulos}}, \bibinfo {author}
  {\bibfnamefont {P.~G.}\ \bibnamefont {Savvidis}}, \ and\ \bibinfo {author}
  {\bibfnamefont {A.}~\bibnamefont {Kavokin}},\ }\href {\doibase
  10.1038/s41598-023-31520-z} {\bibfield  {journal} {\bibinfo  {journal}
  {Scientific Reports}\ }\textbf {\bibinfo {volume} {13}},\ \bibinfo {pages}
  {4607} (\bibinfo {year} {2023})}\BibitemShut {NoStop}%
\bibitem [{\citenamefont {Toebes}\ \emph {et~al.}(2022)\citenamefont {Toebes},
  \citenamefont {Vretenar},\ and\ \citenamefont {Klaers}}]{CommunPhys559}%
  \BibitemOpen
  \bibfield  {author} {\bibinfo {author} {\bibfnamefont {C.}~\bibnamefont
  {Toebes}}, \bibinfo {author} {\bibfnamefont {M.}~\bibnamefont {Vretenar}}, \
  and\ \bibinfo {author} {\bibfnamefont {J.}~\bibnamefont {Klaers}},\ }\href
  {\doibase 10.1038/s42005-022-00832-3} {\bibfield  {journal} {\bibinfo
  {journal} {Communications Physics}\ }\textbf {\bibinfo {volume} {5}},\
  \bibinfo {pages} {59} (\bibinfo {year} {2022})}\BibitemShut {NoStop}%
\bibitem [{\citenamefont {{Pete
  Warden}}(2018)}]{warden2018speechcommandsdatasetlimitedvocabulary}%
  \BibitemOpen
  \bibfield  {author} {\bibinfo {author} {\bibnamefont {{Pete Warden}}},\
  }\href {https://arxiv.org/abs/1804.03209} {\enquote {\bibinfo {title} {Speech
  commands: A dataset for limited-vocabulary speech recognition},}\ } (\bibinfo
  {year} {2018}),\ \Eprint {http://arxiv.org/abs/1804.03209} {arXiv:1804.03209
  [cs.CL]} \BibitemShut {NoStop}%
\bibitem [{spe()}]{speechcommandsv2}%
  \BibitemOpen
  \href@noop {} {}\bibinfo {note} {{ Speech commands dataset version 2.
  [Online]. Available:
  \url{http://download.tensorflow.org/data/speech_commands_v0.02.tar.gz}.
  Released under the Creative Commons BY 4.0 license [Online]. Available:
  \url{https://creativecommons.org/licenses/by/4.0/}}}\BibitemShut {NoStop}%
\bibitem [{\citenamefont {Deng}\ \emph {et~al.}(1991)\citenamefont {Deng},
  \citenamefont {Kenny}, \citenamefont {Lennig}, \citenamefont {Gupta},
  \citenamefont {Seitz},\ and\ \citenamefont {Mermelstein}}]{IEEETrSP134406}%
  \BibitemOpen
  \bibfield  {author} {\bibinfo {author} {\bibfnamefont {L.}~\bibnamefont
  {Deng}}, \bibinfo {author} {\bibfnamefont {P.}~\bibnamefont {Kenny}},
  \bibinfo {author} {\bibfnamefont {M.}~\bibnamefont {Lennig}}, \bibinfo
  {author} {\bibfnamefont {V.}~\bibnamefont {Gupta}}, \bibinfo {author}
  {\bibfnamefont {F.}~\bibnamefont {Seitz}}, \ and\ \bibinfo {author}
  {\bibfnamefont {P.}~\bibnamefont {Mermelstein}},\ }\href {\doibase
  10.1109/78.134406} {\bibfield  {journal} {\bibinfo  {journal} {IEEE
  Transactions on Signal Processing}\ }\textbf {\bibinfo {volume} {39}},\
  \bibinfo {pages} {1677} (\bibinfo {year} {1991})}\BibitemShut {NoStop}%
\bibitem [{\citenamefont {Gales}\ and\ \citenamefont
  {Young}(2008)}]{SIG0042008}%
  \BibitemOpen
  \bibfield  {author} {\bibinfo {author} {\bibfnamefont {M.}~\bibnamefont
  {Gales}}\ and\ \bibinfo {author} {\bibfnamefont {S.}~\bibnamefont {Young}},\
  }\href {\doibase 10.1561/2000000004} {\bibfield  {journal} {\bibinfo
  {journal} {Foundations and Trends® in Signal Processing}\ }\textbf {\bibinfo
  {volume} {1}},\ \bibinfo {pages} {195} (\bibinfo {year} {2008})}\BibitemShut
  {NoStop}%
\bibitem [{\citenamefont {Kasprzak}\ \emph {et~al.}(2006)\citenamefont
  {Kasprzak}, \citenamefont {Richard}, \citenamefont {Kundermann},
  \citenamefont {Baas}, \citenamefont {Jeambrun}, \citenamefont {Keeling},
  \citenamefont {Marchetti}, \citenamefont {Szyma{\'n}ska}, \citenamefont
  {Andr{\'e}}, \citenamefont {Staehli}, \citenamefont {Savona}, \citenamefont
  {Littlewood}, \citenamefont {Deveaud},\ and\ \citenamefont
  {Dang}}]{Nature443409}%
  \BibitemOpen
  \bibfield  {author} {\bibinfo {author} {\bibfnamefont {J.}~\bibnamefont
  {Kasprzak}}, \bibinfo {author} {\bibfnamefont {M.}~\bibnamefont {Richard}},
  \bibinfo {author} {\bibfnamefont {S.}~\bibnamefont {Kundermann}}, \bibinfo
  {author} {\bibfnamefont {A.}~\bibnamefont {Baas}}, \bibinfo {author}
  {\bibfnamefont {P.}~\bibnamefont {Jeambrun}}, \bibinfo {author}
  {\bibfnamefont {J.~M.~J.}\ \bibnamefont {Keeling}}, \bibinfo {author}
  {\bibfnamefont {F.~M.}\ \bibnamefont {Marchetti}}, \bibinfo {author}
  {\bibfnamefont {M.~H.}\ \bibnamefont {Szyma{\'n}ska}}, \bibinfo {author}
  {\bibfnamefont {R.}~\bibnamefont {Andr{\'e}}}, \bibinfo {author}
  {\bibfnamefont {J.~L.}\ \bibnamefont {Staehli}}, \bibinfo {author}
  {\bibfnamefont {V.}~\bibnamefont {Savona}}, \bibinfo {author} {\bibfnamefont
  {P.~B.}\ \bibnamefont {Littlewood}}, \bibinfo {author} {\bibfnamefont
  {B.}~\bibnamefont {Deveaud}}, \ and\ \bibinfo {author} {\bibfnamefont
  {L.~S.}\ \bibnamefont {Dang}},\ }\href {\doibase 10.1038/nature05131}
  {\bibfield  {journal} {\bibinfo  {journal} {Nature}\ }\textbf {\bibinfo
  {volume} {443}},\ \bibinfo {pages} {409} (\bibinfo {year}
  {2006})}\BibitemShut {NoStop}%
\bibitem [{\citenamefont {Sedov}\ \emph {et~al.}(2020)\citenamefont {Sedov},
  \citenamefont {Lukoshkin}, \citenamefont {Kalevich}, \citenamefont
  {Hatzopoulos}, \citenamefont {Savvidis},\ and\ \citenamefont
  {Kavokin}}]{ACSPhot71163}%
  \BibitemOpen
  \bibfield  {author} {\bibinfo {author} {\bibfnamefont {E.}~\bibnamefont
  {Sedov}}, \bibinfo {author} {\bibfnamefont {V.}~\bibnamefont {Lukoshkin}},
  \bibinfo {author} {\bibfnamefont {V.}~\bibnamefont {Kalevich}}, \bibinfo
  {author} {\bibfnamefont {Z.}~\bibnamefont {Hatzopoulos}}, \bibinfo {author}
  {\bibfnamefont {P.}~\bibnamefont {Savvidis}}, \ and\ \bibinfo {author}
  {\bibfnamefont {A.}~\bibnamefont {Kavokin}},\ }\href {\doibase
  10.1021/acsphotonics.9b01779} {\bibfield  {journal} {\bibinfo  {journal} {ACS
  Photonics}\ }\textbf {\bibinfo {volume} {7}},\ \bibinfo {pages} {1163}
  (\bibinfo {year} {2020})}\BibitemShut {NoStop}%
\bibitem [{\citenamefont {Wang}\ \emph {et~al.}(2022)\citenamefont {Wang},
  \citenamefont {Lagoudakis},\ and\ \citenamefont
  {Sigurdsson}}]{PhysRevB106245304}%
  \BibitemOpen
  \bibfield  {author} {\bibinfo {author} {\bibfnamefont {Y.}~\bibnamefont
  {Wang}}, \bibinfo {author} {\bibfnamefont {P.~G.}\ \bibnamefont
  {Lagoudakis}}, \ and\ \bibinfo {author} {\bibfnamefont {H.}~\bibnamefont
  {Sigurdsson}},\ }\href {\doibase 10.1103/PhysRevB.106.245304} {\bibfield
  {journal} {\bibinfo  {journal} {Phys. Rev. B}\ }\textbf {\bibinfo {volume}
  {106}},\ \bibinfo {pages} {245304} (\bibinfo {year} {2022})}\BibitemShut
  {NoStop}%
\bibitem [{\citenamefont {{Davis, S. and Mermelstein,
  P.}}(1980)}]{IEEE1163420}%
  \BibitemOpen
  \bibfield  {author} {\bibinfo {author} {\bibnamefont {{Davis, S. and
  Mermelstein, P.}}},\ }\href {\doibase 10.1109/TASSP.1980.1163420} {\bibfield
  {journal} {\bibinfo  {journal} {IEEE Transactions on Acoustics, Speech, and
  Signal Processing}\ }\textbf {\bibinfo {volume} {28}},\ \bibinfo {pages}
  {357} (\bibinfo {year} {1980})}\BibitemShut {NoStop}%
\bibitem [{\citenamefont {{Md. Sahidullah and Goutam
  Saha}}(2012)}]{SAHIDULLAH2012543}%
  \BibitemOpen
  \bibfield  {author} {\bibinfo {author} {\bibnamefont {{Md. Sahidullah and
  Goutam Saha}}},\ }\href {\doibase
  https://doi.org/10.1016/j.specom.2011.11.004} {\bibfield  {journal} {\bibinfo
   {journal} {Speech Communication}\ }\textbf {\bibinfo {volume} {54}},\
  \bibinfo {pages} {543} (\bibinfo {year} {2012})}\BibitemShut {NoStop}%
\bibitem [{\citenamefont {{Sumedha Rai and Tong Li and Bella
  Lyu}}(2023)}]{rai2023keywordspottingdetecting}%
  \BibitemOpen
  \bibfield  {author} {\bibinfo {author} {\bibnamefont {{Sumedha Rai and Tong
  Li and Bella Lyu}}},\ }\href {https://arxiv.org/abs/2312.05640} {\enquote
  {\bibinfo {title} {Keyword spotting -- detecting commands in speech using
  deep learning},}\ } (\bibinfo {year} {2023}),\ \Eprint
  {http://arxiv.org/abs/2312.05640} {arXiv:2312.05640 [cs.SD]} \BibitemShut
  {NoStop}%
\bibitem [{\citenamefont {Chang}\ \emph {et~al.}(2018)\citenamefont {Chang},
  \citenamefont {Sitzmann}, \citenamefont {Dun}, \citenamefont {Heidrich},\
  and\ \citenamefont {Wetzstein}}]{SciRep812324}%
  \BibitemOpen
  \bibfield  {author} {\bibinfo {author} {\bibfnamefont {J.}~\bibnamefont
  {Chang}}, \bibinfo {author} {\bibfnamefont {V.}~\bibnamefont {Sitzmann}},
  \bibinfo {author} {\bibfnamefont {X.}~\bibnamefont {Dun}}, \bibinfo {author}
  {\bibfnamefont {W.}~\bibnamefont {Heidrich}}, \ and\ \bibinfo {author}
  {\bibfnamefont {G.}~\bibnamefont {Wetzstein}},\ }\href {\doibase
  10.1038/s41598-018-30619-y} {\bibfield  {journal} {\bibinfo  {journal}
  {Scientific Reports}\ }\textbf {\bibinfo {volume} {8}},\ \bibinfo {pages}
  {12324} (\bibinfo {year} {2018})}\BibitemShut {NoStop}%
\bibitem [{\citenamefont {Spall}\ \emph {et~al.}(2020)\citenamefont {Spall},
  \citenamefont {Guo}, \citenamefont {Barrett},\ and\ \citenamefont
  {Lvovsky}}]{OptLett455752}%
  \BibitemOpen
  \bibfield  {author} {\bibinfo {author} {\bibfnamefont {J.}~\bibnamefont
  {Spall}}, \bibinfo {author} {\bibfnamefont {X.}~\bibnamefont {Guo}}, \bibinfo
  {author} {\bibfnamefont {T.~D.}\ \bibnamefont {Barrett}}, \ and\ \bibinfo
  {author} {\bibfnamefont {A.~I.}\ \bibnamefont {Lvovsky}},\ }\href {\doibase
  10.1364/OL.401675} {\bibfield  {journal} {\bibinfo  {journal} {Opt. Lett.}\
  }\textbf {\bibinfo {volume} {45}},\ \bibinfo {pages} {5752} (\bibinfo {year}
  {2020})}\BibitemShut {NoStop}%
\bibitem [{\citenamefont {Ma}\ and\ \citenamefont
  {Schumacher}(2017)}]{PhysRevB95235301}%
  \BibitemOpen
  \bibfield  {author} {\bibinfo {author} {\bibfnamefont {X.}~\bibnamefont
  {Ma}}\ and\ \bibinfo {author} {\bibfnamefont {S.}~\bibnamefont
  {Schumacher}},\ }\href {\doibase 10.1103/PhysRevB.95.235301} {\bibfield
  {journal} {\bibinfo  {journal} {Phys. Rev. B}\ }\textbf {\bibinfo {volume}
  {95}},\ \bibinfo {pages} {235301} (\bibinfo {year} {2017})}\BibitemShut
  {NoStop}%
\bibitem [{\citenamefont {Struck}\ \emph {et~al.}(2011)\citenamefont {Struck},
  \citenamefont {Ölschläger}, \citenamefont {Targat}, \citenamefont
  {Soltan-Panahi}, \citenamefont {Eckardt}, \citenamefont {Lewenstein},
  \citenamefont {Windpassinger},\ and\ \citenamefont
  {Sengstock}}]{Science333996}%
  \BibitemOpen
  \bibfield  {author} {\bibinfo {author} {\bibfnamefont {J.}~\bibnamefont
  {Struck}}, \bibinfo {author} {\bibfnamefont {C.}~\bibnamefont
  {Ölschläger}}, \bibinfo {author} {\bibfnamefont {R.~L.}\ \bibnamefont
  {Targat}}, \bibinfo {author} {\bibfnamefont {P.}~\bibnamefont
  {Soltan-Panahi}}, \bibinfo {author} {\bibfnamefont {A.}~\bibnamefont
  {Eckardt}}, \bibinfo {author} {\bibfnamefont {M.}~\bibnamefont {Lewenstein}},
  \bibinfo {author} {\bibfnamefont {P.}~\bibnamefont {Windpassinger}}, \ and\
  \bibinfo {author} {\bibfnamefont {K.}~\bibnamefont {Sengstock}},\ }\href
  {\doibase 10.1126/science.1207239} {\bibfield  {journal} {\bibinfo  {journal}
  {Science}\ }\textbf {\bibinfo {volume} {333}},\ \bibinfo {pages} {996}
  (\bibinfo {year} {2011})}\BibitemShut {NoStop}%
\bibitem [{\citenamefont {Bloch}\ \emph {et~al.}(2012)\citenamefont {Bloch},
  \citenamefont {Dalibard},\ and\ \citenamefont
  {Nascimb{\`e}ne}}]{NatPhys8267}%
  \BibitemOpen
  \bibfield  {author} {\bibinfo {author} {\bibfnamefont {I.}~\bibnamefont
  {Bloch}}, \bibinfo {author} {\bibfnamefont {J.}~\bibnamefont {Dalibard}}, \
  and\ \bibinfo {author} {\bibfnamefont {S.}~\bibnamefont {Nascimb{\`e}ne}},\
  }\href {\doibase 10.1038/nphys2259} {\bibfield  {journal} {\bibinfo
  {journal} {Nature Physics}\ }\textbf {\bibinfo {volume} {8}},\ \bibinfo
  {pages} {267} (\bibinfo {year} {2012})}\BibitemShut {NoStop}%
\bibitem [{\citenamefont {Gao}\ \emph {et~al.}(2012)\citenamefont {Gao},
  \citenamefont {Eldridge}, \citenamefont {Liew}, \citenamefont {Tsintzos},
  \citenamefont {Stavrinidis}, \citenamefont {Deligeorgis}, \citenamefont
  {Hatzopoulos},\ and\ \citenamefont {Savvidis}}]{PhysRevB85235102}%
  \BibitemOpen
  \bibfield  {author} {\bibinfo {author} {\bibfnamefont {T.}~\bibnamefont
  {Gao}}, \bibinfo {author} {\bibfnamefont {P.~S.}\ \bibnamefont {Eldridge}},
  \bibinfo {author} {\bibfnamefont {T.~C.~H.}\ \bibnamefont {Liew}}, \bibinfo
  {author} {\bibfnamefont {S.~I.}\ \bibnamefont {Tsintzos}}, \bibinfo {author}
  {\bibfnamefont {G.}~\bibnamefont {Stavrinidis}}, \bibinfo {author}
  {\bibfnamefont {G.}~\bibnamefont {Deligeorgis}}, \bibinfo {author}
  {\bibfnamefont {Z.}~\bibnamefont {Hatzopoulos}}, \ and\ \bibinfo {author}
  {\bibfnamefont {P.~G.}\ \bibnamefont {Savvidis}},\ }\href {\doibase
  10.1103/PhysRevB.85.235102} {\bibfield  {journal} {\bibinfo  {journal} {Phys.
  Rev. B}\ }\textbf {\bibinfo {volume} {85}},\ \bibinfo {pages} {235102}
  (\bibinfo {year} {2012})}\BibitemShut {NoStop}%
\bibitem [{\citenamefont {Schmidt}\ \emph {et~al.}(2019)\citenamefont
  {Schmidt}, \citenamefont {Berger}, \citenamefont {Kahlert}, \citenamefont
  {Bayer}, \citenamefont {Schneider}, \citenamefont {H\"ofling}, \citenamefont
  {Sedov}, \citenamefont {Kavokin},\ and\ \citenamefont
  {A\ss{}mann}}]{PhysRevLett122047403}%
  \BibitemOpen
  \bibfield  {author} {\bibinfo {author} {\bibfnamefont {D.}~\bibnamefont
  {Schmidt}}, \bibinfo {author} {\bibfnamefont {B.}~\bibnamefont {Berger}},
  \bibinfo {author} {\bibfnamefont {M.}~\bibnamefont {Kahlert}}, \bibinfo
  {author} {\bibfnamefont {M.}~\bibnamefont {Bayer}}, \bibinfo {author}
  {\bibfnamefont {C.}~\bibnamefont {Schneider}}, \bibinfo {author}
  {\bibfnamefont {S.}~\bibnamefont {H\"ofling}}, \bibinfo {author}
  {\bibfnamefont {E.~S.}\ \bibnamefont {Sedov}}, \bibinfo {author}
  {\bibfnamefont {A.~V.}\ \bibnamefont {Kavokin}}, \ and\ \bibinfo {author}
  {\bibfnamefont {M.}~\bibnamefont {A\ss{}mann}},\ }\href {\doibase
  10.1103/PhysRevLett.122.047403} {\bibfield  {journal} {\bibinfo  {journal}
  {Phys. Rev. Lett.}\ }\textbf {\bibinfo {volume} {122}},\ \bibinfo {pages}
  {047403} (\bibinfo {year} {2019})}\BibitemShut {NoStop}%
\bibitem [{\citenamefont {Rozas}\ \emph {et~al.}(2023)\citenamefont {Rozas},
  \citenamefont {Sedov}, \citenamefont {Brune}, \citenamefont {H\"ofling},
  \citenamefont {Kavokin},\ and\ \citenamefont
  {A\ss{}mann}}]{PhysRevB108165411}%
  \BibitemOpen
  \bibfield  {author} {\bibinfo {author} {\bibfnamefont {E.}~\bibnamefont
  {Rozas}}, \bibinfo {author} {\bibfnamefont {E.}~\bibnamefont {Sedov}},
  \bibinfo {author} {\bibfnamefont {Y.}~\bibnamefont {Brune}}, \bibinfo
  {author} {\bibfnamefont {S.}~\bibnamefont {H\"ofling}}, \bibinfo {author}
  {\bibfnamefont {A.}~\bibnamefont {Kavokin}}, \ and\ \bibinfo {author}
  {\bibfnamefont {M.}~\bibnamefont {A\ss{}mann}},\ }\href {\doibase
  10.1103/PhysRevB.108.165411} {\bibfield  {journal} {\bibinfo  {journal}
  {Phys. Rev. B}\ }\textbf {\bibinfo {volume} {108}},\ \bibinfo {pages}
  {165411} (\bibinfo {year} {2023})}\BibitemShut {NoStop}%
\bibitem [{\citenamefont {Kalinin}\ and\ \citenamefont
  {Berloff}(2020)}]{AdvQT31900065}%
  \BibitemOpen
  \bibfield  {author} {\bibinfo {author} {\bibfnamefont {K.~P.}\ \bibnamefont
  {Kalinin}}\ and\ \bibinfo {author} {\bibfnamefont {N.~G.}\ \bibnamefont
  {Berloff}},\ }\href {\doibase https://doi.org/10.1002/qute.201900065}
  {\bibfield  {journal} {\bibinfo  {journal} {Advanced Quantum Technologies}\
  }\textbf {\bibinfo {volume} {3}},\ \bibinfo {pages} {1900065} (\bibinfo
  {year} {2020})}\BibitemShut {NoStop}%
\bibitem [{\citenamefont {Schneider}\ \emph {et~al.}(2016)\citenamefont
  {Schneider}, \citenamefont {Winkler}, \citenamefont {Fraser}, \citenamefont
  {Kamp}, \citenamefont {Yamamoto}, \citenamefont {Ostrovskaya},\ and\
  \citenamefont {Höfling}}]{RepProgrPhys80016503}%
  \BibitemOpen
  \bibfield  {author} {\bibinfo {author} {\bibfnamefont {C.}~\bibnamefont
  {Schneider}}, \bibinfo {author} {\bibfnamefont {K.}~\bibnamefont {Winkler}},
  \bibinfo {author} {\bibfnamefont {M.~D.}\ \bibnamefont {Fraser}}, \bibinfo
  {author} {\bibfnamefont {M.}~\bibnamefont {Kamp}}, \bibinfo {author}
  {\bibfnamefont {Y.}~\bibnamefont {Yamamoto}}, \bibinfo {author}
  {\bibfnamefont {E.~A.}\ \bibnamefont {Ostrovskaya}}, \ and\ \bibinfo {author}
  {\bibfnamefont {S.}~\bibnamefont {Höfling}},\ }\href {\doibase
  10.1088/0034-4885/80/1/016503} {\bibfield  {journal} {\bibinfo  {journal}
  {Reports on Progress in Physics}\ }\textbf {\bibinfo {volume} {80}},\
  \bibinfo {pages} {016503} (\bibinfo {year} {2016})}\BibitemShut {NoStop}%
\bibitem [{\citenamefont {Balili}\ \emph {et~al.}(2007)\citenamefont {Balili},
  \citenamefont {Hartwell}, \citenamefont {Snoke}, \citenamefont {Pfeiffer},\
  and\ \citenamefont {West}}]{Science3161007}%
  \BibitemOpen
  \bibfield  {author} {\bibinfo {author} {\bibfnamefont {R.}~\bibnamefont
  {Balili}}, \bibinfo {author} {\bibfnamefont {V.}~\bibnamefont {Hartwell}},
  \bibinfo {author} {\bibfnamefont {D.}~\bibnamefont {Snoke}}, \bibinfo
  {author} {\bibfnamefont {L.}~\bibnamefont {Pfeiffer}}, \ and\ \bibinfo
  {author} {\bibfnamefont {K.}~\bibnamefont {West}},\ }\href {\doibase
  10.1126/science.1140990} {\bibfield  {journal} {\bibinfo  {journal}
  {Science}\ }\textbf {\bibinfo {volume} {316}},\ \bibinfo {pages} {1007}
  (\bibinfo {year} {2007})}\BibitemShut {NoStop}%
\bibitem [{\citenamefont {Sanvitto}\ and\ \citenamefont
  {K{\'e}na-Cohen}(2016)}]{NatMat151061}%
  \BibitemOpen
  \bibfield  {author} {\bibinfo {author} {\bibfnamefont {D.}~\bibnamefont
  {Sanvitto}}\ and\ \bibinfo {author} {\bibfnamefont {S.}~\bibnamefont
  {K{\'e}na-Cohen}},\ }\href {\doibase 10.1038/nmat4668} {\bibfield  {journal}
  {\bibinfo  {journal} {Nature Materials}\ }\textbf {\bibinfo {volume} {15}},\
  \bibinfo {pages} {1061} (\bibinfo {year} {2016})}\BibitemShut {NoStop}%
\bibitem [{\citenamefont {Borri}\ \emph {et~al.}(2000)\citenamefont {Borri},
  \citenamefont {Langbein}, \citenamefont {Woggon}, \citenamefont {Jensen},\
  and\ \citenamefont {Hvam}}]{PhysRevB62R7763}%
  \BibitemOpen
  \bibfield  {author} {\bibinfo {author} {\bibfnamefont {P.}~\bibnamefont
  {Borri}}, \bibinfo {author} {\bibfnamefont {W.}~\bibnamefont {Langbein}},
  \bibinfo {author} {\bibfnamefont {U.}~\bibnamefont {Woggon}}, \bibinfo
  {author} {\bibfnamefont {J.~R.}\ \bibnamefont {Jensen}}, \ and\ \bibinfo
  {author} {\bibfnamefont {J.~M.}\ \bibnamefont {Hvam}},\ }\href {\doibase
  10.1103/PhysRevB.62.R7763} {\bibfield  {journal} {\bibinfo  {journal} {Phys.
  Rev. B}\ }\textbf {\bibinfo {volume} {62}},\ \bibinfo {pages} {R7763}
  (\bibinfo {year} {2000})}\BibitemShut {NoStop}%
\bibitem [{\citenamefont {Galbiati}\ \emph {et~al.}(2012)\citenamefont
  {Galbiati}, \citenamefont {Ferrier}, \citenamefont {Solnyshkov},
  \citenamefont {Tanese}, \citenamefont {Wertz}, \citenamefont {Amo},
  \citenamefont {Abbarchi}, \citenamefont {Senellart}, \citenamefont {Sagnes},
  \citenamefont {Lema\^{\i}tre}, \citenamefont {Galopin}, \citenamefont
  {Malpuech},\ and\ \citenamefont {Bloch}}]{PhysRevLett108126403}%
  \BibitemOpen
  \bibfield  {author} {\bibinfo {author} {\bibfnamefont {M.}~\bibnamefont
  {Galbiati}}, \bibinfo {author} {\bibfnamefont {L.}~\bibnamefont {Ferrier}},
  \bibinfo {author} {\bibfnamefont {D.~D.}\ \bibnamefont {Solnyshkov}},
  \bibinfo {author} {\bibfnamefont {D.}~\bibnamefont {Tanese}}, \bibinfo
  {author} {\bibfnamefont {E.}~\bibnamefont {Wertz}}, \bibinfo {author}
  {\bibfnamefont {A.}~\bibnamefont {Amo}}, \bibinfo {author} {\bibfnamefont
  {M.}~\bibnamefont {Abbarchi}}, \bibinfo {author} {\bibfnamefont
  {P.}~\bibnamefont {Senellart}}, \bibinfo {author} {\bibfnamefont
  {I.}~\bibnamefont {Sagnes}}, \bibinfo {author} {\bibfnamefont
  {A.}~\bibnamefont {Lema\^{\i}tre}}, \bibinfo {author} {\bibfnamefont
  {E.}~\bibnamefont {Galopin}}, \bibinfo {author} {\bibfnamefont
  {G.}~\bibnamefont {Malpuech}}, \ and\ \bibinfo {author} {\bibfnamefont
  {J.}~\bibnamefont {Bloch}},\ }\href {\doibase 10.1103/PhysRevLett.108.126403}
  {\bibfield  {journal} {\bibinfo  {journal} {Phys. Rev. Lett.}\ }\textbf
  {\bibinfo {volume} {108}},\ \bibinfo {pages} {126403} (\bibinfo {year}
  {2012})}\BibitemShut {NoStop}%
\bibitem [{\citenamefont {Baboux}\ \emph {et~al.}(2016)\citenamefont {Baboux},
  \citenamefont {Ge}, \citenamefont {Jacqmin}, \citenamefont {Biondi},
  \citenamefont {Galopin}, \citenamefont {Lema\^{\i}tre}, \citenamefont
  {Le~Gratiet}, \citenamefont {Sagnes}, \citenamefont {Schmidt}, \citenamefont
  {T\"ureci}, \citenamefont {Amo},\ and\ \citenamefont
  {Bloch}}]{PhysRevLett116066402}%
  \BibitemOpen
  \bibfield  {author} {\bibinfo {author} {\bibfnamefont {F.}~\bibnamefont
  {Baboux}}, \bibinfo {author} {\bibfnamefont {L.}~\bibnamefont {Ge}}, \bibinfo
  {author} {\bibfnamefont {T.}~\bibnamefont {Jacqmin}}, \bibinfo {author}
  {\bibfnamefont {M.}~\bibnamefont {Biondi}}, \bibinfo {author} {\bibfnamefont
  {E.}~\bibnamefont {Galopin}}, \bibinfo {author} {\bibfnamefont
  {A.}~\bibnamefont {Lema\^{\i}tre}}, \bibinfo {author} {\bibfnamefont
  {L.}~\bibnamefont {Le~Gratiet}}, \bibinfo {author} {\bibfnamefont
  {I.}~\bibnamefont {Sagnes}}, \bibinfo {author} {\bibfnamefont
  {S.}~\bibnamefont {Schmidt}}, \bibinfo {author} {\bibfnamefont {H.~E.}\
  \bibnamefont {T\"ureci}}, \bibinfo {author} {\bibfnamefont {A.}~\bibnamefont
  {Amo}}, \ and\ \bibinfo {author} {\bibfnamefont {J.}~\bibnamefont {Bloch}},\
  }\href {\doibase 10.1103/PhysRevLett.116.066402} {\bibfield  {journal}
  {\bibinfo  {journal} {Phys. Rev. Lett.}\ }\textbf {\bibinfo {volume} {116}},\
  \bibinfo {pages} {066402} (\bibinfo {year} {2016})}\BibitemShut {NoStop}%
\bibitem [{\citenamefont {Kurtscheid}\ \emph {et~al.}(2020)\citenamefont
  {Kurtscheid}, \citenamefont {Dung}, \citenamefont {Redmann}, \citenamefont
  {Busley}, \citenamefont {Klaers}, \citenamefont {Vewinger}, \citenamefont
  {Schmitt},\ and\ \citenamefont {Weitz}}]{EPL13054001}%
  \BibitemOpen
  \bibfield  {author} {\bibinfo {author} {\bibfnamefont {C.}~\bibnamefont
  {Kurtscheid}}, \bibinfo {author} {\bibfnamefont {D.}~\bibnamefont {Dung}},
  \bibinfo {author} {\bibfnamefont {A.}~\bibnamefont {Redmann}}, \bibinfo
  {author} {\bibfnamefont {E.}~\bibnamefont {Busley}}, \bibinfo {author}
  {\bibfnamefont {J.}~\bibnamefont {Klaers}}, \bibinfo {author} {\bibfnamefont
  {F.}~\bibnamefont {Vewinger}}, \bibinfo {author} {\bibfnamefont
  {J.}~\bibnamefont {Schmitt}}, \ and\ \bibinfo {author} {\bibfnamefont
  {M.}~\bibnamefont {Weitz}},\ }\href {\doibase 10.1209/0295-5075/130/54001}
  {\bibfield  {journal} {\bibinfo  {journal} {Europhysics Letters}\ }\textbf
  {\bibinfo {volume} {130}},\ \bibinfo {pages} {54001} (\bibinfo {year}
  {2020})}\BibitemShut {NoStop}%
\end{thebibliography}%

%+++++++++++++++++++++++++
%+++++++++++++++++++++++++
%+++++++++++++++++++++++++
\section*{Acknowledgments}
The support of Saint-Petersburg State University (research grant No.~122040800257-5),
the state assignment in the field of scientific activity of the Ministry of Science and Higher Education of the Russian Federation (theme FZUN-2024-0019, state assignment of VlSU) and the Innovation Program for Quantum Science and Technology 2023ZD0300300 are acknowledged.

\section*{Authors' contribution}
E.S. devised the concept, performed simulations and analyzed the simulation results.
A.K. initiated the study and guided the work.
Both authors wrote the manuscript, discussed and contributed to the paper.

\section*{Conflict of interest}
The authors declare that they have no conflict of interest.

\section*{Data availability}
All data generated and analyzed during the current study are available from the corresponding author on reasonable request.

\end{document}